\documentclass[prb,showpacs]{revtex4}
\usepackage{graphicx,amssymb,bm,makeidx}

\usepackage{amssymb}
\usepackage{amsmath}
\usepackage{grffile}

\newcommand{\kk}{\mathbf{k}}
\newcommand{\I}{\mathrm{i}}
\newcommand{\nVH}{n_{\rm vH}}

\begin{document}


\title{Spiral magnetism in the single-band Hubbard model: the Hartree--Fock and slave-boson approaches}

\author{P.A. Igoshev}
\email{igoshev_pa@imp.uran.ru}
\affiliation{Institute of Metal Physics, 620990, Kovalevskaya str. 18, Ekaterinburg, Russia}
\affiliation{Ural Federal University, 620002 Ekaterinburg, Russia}

\author{M.A. Timirgazin}
\author{V.F. Gilmutdinov}
\author{A.K. Arzhnikov}
\affiliation{Physical-Technical Institute, 426000, Kirov str. 132, Izhevsk, Russia}

\author{V.Yu. Irkhin}
\affiliation{Institute of Metal Physics, 620990, Kovalevskaya str. 18, Ekaterinburg, Russia}




\date{\today}

\begin{abstract}
The ground-state magnetic phase diagram is investigated within the single-band Hubbard model for square and different cubic lattices. The results of employing  the generalized non-correlated mean-field (Hartree--Fock) approximation and generalized slave-boson approach by Kotliar and Ruckenstein with correlation effects included are compared. We take into account commensurate ferromagnetic, antiferromagnetic, and incommensurate (spiral) magnetic phases, as well as phase separation into magnetic phases of different types, which was often lacking in previous investigations. It is found  that the spiral states and especially ferromagnetism are generally strongly suppressed up to non-realistically large Hubbard $U$  by the correlation effects if nesting is absent and van Hove singularities are well away from the paramagnetic phase Fermi level. The magnetic phase separation plays an important role in the formation of magnetic states, the corresponding phase regions being especially wide in the vicinity of half-filling. The details of non-collinear and collinear magnetic ordering for different cubic lattices are discussed.
\end{abstract}


\maketitle

\section{Introduction}
\label{sec:intro}

Magnetic properties of strongly correlated transition-metal compounds and their relation to doping, lattice geometry,
band structure and interaction parameters are still being extensively investigated \cite{Imada98,Penn}. 
In particular, the details of magnetic order in the ground state remain to be examined both theoretically and experimentally.
During recent decades, the two-dimensional (2D) case closely related to the problem of high-temperature
superconductivity in cuprates and iron arsenides has been intensively investigated theoretically. However, the more
widespread case of three-dimensional compounds is not yet sufficiently understood.
In general, the following factors should be taken into account to state an appropriate physical picture of the magnetism formation:
(i) spiral magnetism, (ii) phase separation, (iii) electron correlation impact, (iv) multiorbital effects.

The single-band Hubbard model provides a sufficient basis for an analysis of complex magnetic properties.
In the case of a large Coulomb interaction parameter the ground state of single-band model in the half-filled case
for bipartite lattices is a N\'eel antiferromagnetic (AFM) insulator.
The types of the antiferromagnetic state instability in the presence of doping or the finite integral of the electron transfer between the second neighbors are still incompletely revealed.
According to the classical work by Nagaoka\cite{Nagaoka66}, on addition of one excess charge carrier the ground state on the
bipartite lattice becomes  saturated ferromagnetic (FM).
This statement can be also considered as a reasonable hypothesis in the case of small finite doping~\cite{Nagaoka66,Linden91,Irkhin04}.
At the same time, for the non-bipartite (e.g., fcc and hcp) lattices  the ground state can be more complicated
\cite{Nagaoka66,Iordansky}.
For finite doping, the ferromagnetic state can become non-saturated \cite{Roth,Park,Irkhin04}.

Scenarios of the possible doping-induced magnetic ordering include the phase separation (PS) of different types: into
the
ferromagnetic and antiferromagnetic phases~\cite{Visscher73} or the phase of superconducting electron liquid and the
N\'eel antiferromagnetic phase~\cite{Emery90}. An alternative scenario is the formation of a spiral (incommensurate) magnetic state.
It was considered within different approaches: the analysis of the momentum dependence of the generalized static magnetic susceptibility for the bare spectrum~\cite{Schultz90}, the Hartree--Fock approximation (HFA) (with small and moderate $U/W$ values considered, where $U$ is the Coulomb repulsion parameter and $W$ is the bandwidth)~\cite{Arrigoni91,Igoshev10} and the $t-J$ model (large $U/W$ values)~\cite{Shraiman90,Read89}. 
It should be stressed that within simple HFA-like approaches the paramagnetic (PM) state is, as a rule, less favorable
than the magnetic ones, which seems unreasonable because the PM phase is dominant in the ground state for the
majority of transition metal compounds. 
Another issue is investigating competition of the collinear, incommensurate and commensurate (FM and AFM) phases. 

Non--uniform phase-separated states are observed in both 2D and 3D magnetic systems \cite{Nagaev,Dagotto}.  
The experimental observation of antiferromagnetic or spiral magnetic structures makes the classical problem of theoretical description of the formation of different types of magnetic order very urgent.  
Spiral structures are observed in both two- and three-dimensional compounds:
in iron-based high-temperature superconductors~\cite{Bao09} and fcc iron
precipitates~\cite{Tsunoda}, MnSi systems~\cite{MnSi}, itinerant compounds Y$_2$Fe$_{17}$ and
Lu$_2$Fe$_{17}$\cite{Kamarad,Prokhnenko} where helical magnetism is induced by pressure. 
Incommensurate  structures are found in AFM chromium~\cite{Fawcett88}. 
In doped cuprates such states are observed as the dynamic magnetic order~\cite{HTSCrewiew}.
Besides, considerably enhanced incommensurate magnetic fluctuations are observed in strontium ruthenates at low temperatures~\cite{Ru} (see discussion in Refs.~\cite{Igoshev10,Igoshev07,Igoshev11}).

Although spiral states were widely theoretically studied,  the relation between the type of magnetic order and the model parameters has not been still clearly determined. 
In addition, there is a question about the possibility of the inhomogeneous state formation in a broad sense: either via magnetic phase separation or via  so-called stripes. The latter structures, representing a mixed charge-spin order, are widely discussed for 2D high-$T_{\rm c}$ superconducting systems \cite{Kivelson03,Vedyaev,Machida1990}.

A detailed study of the magnetic phase diagram of the 2D Hubbard model taking into account the electron transfer integral only
between the nearest neighbors, $t$, within HFA shows that the spiral magnetic states occur in a wide range of parameters,
especially at moderate values of $U\lesssim W$~\cite{Sarker91,Dzierzawa92}.
It was shown in Ref. \onlinecite{Igoshev10} that inclusion of the electron transfer between the second neighbors, $t'\ne0$, in the Hamiltonian changes considerably the ground-state magnetic phase diagram. 
It was established that the inhomogeneous stripe structures  can be more energetically favorable than the spiral states at $t'\ne0$ 
in the slave-boson approximation~\cite{Oles}, as well as at $t' = 0$ and $U\leq W$ in the
Hartree--Fock approximation~\cite{Timirgazin12}.
However, this conclusion is somewhat devaluated by disregarding the intersite Coulomb interaction which can considerably increase the energy of the stripes.

Last decades, the dynamical mean--field theory (DMFT) has been intensively developed and became a powerful tool for the solution of quantum many--body problems \cite{Kotliar_DMFT_review}. 
Being well justified in the limit of large space dimensionality $d$, it takes into account local correlations exactly, which can, in principle, provide some unbiased solution of the problem of
ground magnetic state in the Hubbard model. 
In Refs.~\onlinecite{Fleck1998,Fleck1999} a DMFT generalization to spiral magnetic states was proposed. 
A typical HFA pattern of phase transitions between commensurate and incommensurate magnetic states with doping was reproduced for the square lattice. A novel so-called inhomogeneous DMFT based approach
was proposed in Ref.~\onlinecite{Peters2014} to study magnetic state formation for a rather large cluster. The collinear stripe-like spin-density wave states were found at finite doping and the magnetic phase diagram for the square lattice was constructed.

Another convenient approach to study the magnetic order formation with account of correlations is the slave-boson
approximation (SBA) by Kotliar and Ruckenstein~\cite{Kotliar86} within the Hubbard model.
It yields a simple (in computational sense) method taking into account the crucial difference between the single and
double site states on the mean-field level and allows a thermodynamical description of the magnetic phase transition
physics.
In the saddle point approximation, this method is qualitatively similar to the Gutzwiller approximation~\cite{Gutzwiller}.
The ground state  energy obtained within the slave-boson method is in a good agreement with the quantum
Monte-Carlo and exact diagonalization  calculations~\cite{Fresard92}.
This justifies the use of the slave-boson method in the cases where there are no strong fluctuations of the order
parameter in the system.
In the opposite case of the continuous (second-order) phase transition, where strong fluctuations occur, the nontrivial many-particle renormalizations of the Green's functions can be important~\cite{Igoshev07,Katanin03}.

The effect of electron correlations on the stability of  spiral magnetic states was considered in~\cite{Fresard92} using the
SBA by Kotliar and Ruckenstein.
The phase diagram of the Hubbard model was built in the nearest-neighbor approximation  ($t' = 0$).
Derivation of the static magnetic susceptibility in SBA \cite{Li91} allowed one to obtain a criterion for the
PM state instability with respect to the second-order transition into an incommensurate magnetic state. This
generalizes the analogous criterion obtained within the random phase approximation~\cite{Moriya} to the strongly
correlated states. Using the generalized criterion, the finite $t'$ case was studied and a considerable tendency to
FM ordering at hole doping and large $t'/t$ values was found~\cite{Fresard98}. However, the phase transitions
between magnetically ordered states, as well as all first-order transitions cannot be studied within this approach.

The slave-boson approach can be generalized to take into account the multi--orbital local interaction part of the
Hamiltonian with an indirect connection between the corresponding eigenstates and slave-bosons
states~\cite{Lechermann2007}. The method proposed (the so-called spin-rotation-invariant SBA) yields the effective
momentum dependence of the Green's function residue which is missing in the traditional slave-boson approaches.
Recently, a successful combination of the slave-boson approximation and the local density approximation (LDA) was used
for the first-principle investigations of real compounds~\cite{Lechermann_Ru,Lechermann2012} as a numerically
inexpensive way for qualitatively correct treatment of the electron correlation in the electron spectrum beyond LDA.
Further investigations generalized the method for the study of superconductivity~\cite{Isidori2009}, the physical picture being similar to that obtained within multiorbital LDA+DMFT approach. 
The excitation spectrum in the 3D case differs essentially from the 2D case, and finite--temperature behavior can be hardly described by SB method. 
However, in the ground state the SB approach seems to be useful. 
To generalize the SBA to the case of finite temperatures the  Gaussian fluctuations can be included, e.g.~in the framework of statistically consistent Gutzwiller approximation (SGA) \cite{Spalek}.


In the present paper we apply both the non-correlated mean-field (Hartree--Fock) approximation and the slave-boson
approximation to the magnetic ordering of different types (ferromagnetic, antiferromagnetic and spiral) within 2D and 3D Hubbard model and compare the results.
In Section~\ref{sec:method} we present the model and approximations used.  
The main physical factors which favor the spiral structures formation are discussed in
Section ~\ref{sec:nesting}.
Sections \ref{sec:results_2D_lattices} and \ref{sec:results_3D_lattices} are
devoted to the discussion of the results obtained for the 2D and 3D case, respectively.
Conclusions are presented in Section \ref{sec:conclusions}.

\section{Model and slave-boson approach}

\label{sec:method}
We consider the Hubbard model
\begin{equation}
      \label{eq:original_H}
      \mathcal{H}=\sum_{ij\sigma\sigma'} t_{ij}\delta_{\sigma\sigma'} c^\dag_{i\sigma}c^{}_{j\sigma'}+U\sum_i n_{i\uparrow}n_{i\downarrow},
\end{equation}
where the matrix elements of the electron transfer are $t_{ij} = -t$ for the nearest neighbors and $t'$  for the
next-nearest neighbors (we assume $t>0$), $c^\dag_{i\sigma},c^{}_{i\sigma}$ are the electron creation and annihilation
operators,respectively, $n_{i\sigma}=c^\dag_{i\sigma}c^{}_{i\sigma}$, $i$ is the site number, $\sigma$ is the spin projection.

The local spin space rotation around $x$ axis, matching different site magnetization vectors along, say, $z$ axis,  by the angle $\mathbf{QR}_i$ (where $\bf
Q$ is a spiral wave vector, ${\bf R}_i$ is the site position) is applied for the consideration of magnetic spirals. This
maps the spiral magnetic state into an effective ferromagnetic one, but the hopping term in the Hamiltonian becomes
non-diagonal with respect to index $\sigma$: $t_{ij}\delta_{\sigma\sigma'}\rightarrow t^{\sigma\sigma'}_{ij}$ in Eq.
(\ref{eq:original_H}).
\begin{equation}
      \label{eq:original_H_spiral}
      \tilde{\mathcal{H}}=\sum_{ij\sigma\sigma'} t^{\sigma\sigma'}_{ij} c^\dag_{i\sigma}c^{}_{j\sigma'}+U\sum_i n_{i\uparrow}n_{i\downarrow},
\end{equation}
where $t_{ij}^{\sigma\sigma'}=\exp[\mathrm{i}\mathbf{Q}(\mathbf{R}_i-\mathbf{R}_j)\sigma^x]_{\sigma\sigma'}t_{ij}$. 

Further Hartree--Fock treatment of the many--particle Coulomb interaction term replaces it by the one-electron
interaction term with some effective field $U\langle n_{i\bar\sigma}\rangle $ which however mixes the averaged
contributions from singly and doubly occupied states   ($\langle n_{i\sigma}\rangle$ is the statistical averaged
electron density at site $i$ and spin projection $\sigma$).
This is not correct even qualitatively, especially at large $U$.

Since doubly occupied site states are energetically unfavorable, a simple way of taking into account the correlation
effects is to introduce the slave-boson operators $e_i(e_i^\dag)$ , $p_{i\sigma}(p_{i\sigma}^\dag),
d_i(d_i^\dag)$ for empty, singly and doubly occupied states, respectively\cite{Kotliar86}. 
The transitions between the site states originating from intersite electron transfer can be formulated in alternative
terms, but {\it simultaneously} with the conventional Slater determinant based formalism  (related to the electron
creation/annihilation operators).
Conceptually this is close to the Hubbard's $X$-operator formalism\cite{Hubbard2,Irkhin04}, where, however, the site
transition $X$-operators are introduced {\it instead of} the original one-electron operators.
We use the action formalism and pass to the extended Hamiltonian doing in the fermion-boson Fock space, the Coulomb interaction term becoming diagonal with respect to the boson operators (see method details in Ref.~\onlinecite{Kotliar86}):  
\begin{equation}
      \label{eq:H_ext}
      \mathcal{H}_{\rm ext}=\sum_{ij\sigma\sigma'} t^{\sigma\sigma'}_{ij} c^\dag_{i\sigma}c^{}_{j\sigma'}
z^\dag_{i\sigma}z^{}_{j\sigma'}
+U\sum_i d^\dag_{i}d^{}_{i},
\end{equation}
where
\begin{equation}
z_{i\sigma}=(1-d_i^\dag d^{}_i-p^\dag_{i\sigma}p^{}_{i\sigma})^{-1/2}(e^\dag_ip^{}_{i\sigma}+p^\dag_{i\bar\sigma} d^{}_i)(1-e_i^\dag e^{}_i-p^\dag_{i\bar\sigma}p^{}_{i\bar\sigma})^{-1/2}.
\end{equation}
Equation~(\ref{eq:H_ext}) is equivalent to Eq.~(\ref{eq:original_H}) provided that the unphysical site states (which
have no counterparts for the original electron system) are excluded (Eq.~(\ref{eq:eta-constraint}) below) and the
electronic and bosonic subsystems have a strict mutual correspondence (Eq. (\ref{eq:lmb-constraint}) below).
The physical subsector of the domain of operator $\mathcal{H}_{\rm ext}$ is restricted by the following constraints:
\begin{equation}
      \label{eq:eta-constraint}
      e^\dag_ie^{}_i+\sum_\sigma p^\dag_{i\sigma}p^{}_{i\sigma}+d^\dag_id^{}_i=1,
\end{equation}
\begin{equation}
      \label{eq:lmb-constraint}
      d^\dag_id^{}_i+p^\dag_{i\sigma}p^{}_{i\sigma}=c^\dag_{i\sigma}c^{}_{i\sigma}.
\end{equation}

The presence of the constraints can be taken into account within the functional integral formalism through the Lagrange
multipliers (the local bosonic ``chemical potential'' $\eta_i$ for Eq.~(\ref{eq:eta-constraint}) and the on-site
electron--boson ``force'' $\lambda_{i\sigma}$ for Eq.~(\ref{eq:lmb-constraint})) introduced into the action. To this
point the transformation considered has been rigorous, but further exact functional treatment of both bosonic and fermionic fields is hardly possible.
A reasonable physical picture (different from the Hartree--Fock one) at arbitrary $U/t$  can be obtained within the
saddle-point approximation for the bosonic fields and Lagrange multipliers: for the action $\mathcal{S}$ generated by
Eq.~(\ref{eq:H_ext})
we replace the bosonic fields by their extreme {\it real} values which are assumed to be site- and time-independent:
$e^\dag_i,e^{}_i\rightarrow e;\; p^\dag_{i\sigma},p^{}_{i\sigma}\rightarrow p^{}_{\sigma};\; d^\dag_i,d^{}_i\rightarrow
d;\; \eta_i\rightarrow \eta,\; \lambda_{i\sigma}\rightarrow \lambda_{\sigma}$. 
Explicit equations for the extreme values are presented below.
Note that within this approximation the numbers $e,p_\sigma,d$ have a simple meaning: these are the probability
amplitudes of any site's being
 in the corresponding many-electron state.

The fermionic part of the action can be produced by the Hamiltonian
\begin{equation}
	\mathcal{H}_{\rm f} = \sum_{\sigma\sigma'ij} (z_\sigma z_{\sigma'}t_{ij}^{\sigma\sigma'}+\delta_{ij}\delta_{\sigma\sigma'}\lambda_\sigma) c^\dag_{i\sigma}c^{}_{j\sigma'},
\end{equation}
the corresponding eigenvalues (antiferromagnetic subbands) yielding the renormalization of the electron spectrum:
\begin{equation}
	\label{eq:subband_spectrum}
E_{s=\pm1}(\kk)=(1/2)\left((z^2_\uparrow+z^2_\downarrow)e^+_{\kk}+\lambda_\uparrow+\lambda_\downarrow\right)+ s\sqrt{D_\kk},
\end{equation}
where
\begin{equation}
    \label{eq:Dk}
    D_\kk=(1/4)\left((z^2_\uparrow-z^2_\downarrow)e^+_{\kk}+ \lambda_\uparrow-\lambda_\downarrow\right)^2+(e^-_{\kk}z_\uparrow z_\downarrow)^2
\end{equation}
and
\begin{equation}
    \label{eq:esa}
    e^\pm_{\kk} =(1/2)(t_{\kk+\mathbf{Q}/2}\pm t_{\kk-\mathbf{Q}/2}),
\end{equation}
\begin{equation}
t_\kk = \sum_{i}\exp(\I\kk(\mathbf{R}_i-\mathbf{R}_j))t_{ij}
\end{equation}
is the bare electron spectrum. The subband narrowing is the $c$-number function of extremal bosonic fields
\begin{equation}
	\label{eq:z_def2}
	z^2_\sigma=(1-d^2-p_\sigma^2) (ep_\sigma+p_{\bar\sigma}d)^2 (1-e^2-p_{\bar\sigma}^2).
\end{equation}

The impact of the averaged site states on electron states manifests itself in two types of bare spectrum
renormalization: (i) narrowing of the bare spectrum  (similar to the Hubbard--I approximation\cite{Hubbard}) which is
specified by the factor $z^2_\sigma\le 1$, (ii) the additional energy shift $\lambda_\sigma$.
Note that within HFA
\begin{equation}
\lambda_{\sigma}\rightarrow\lambda^{\rm HF}_{\sigma}=Un/2-Um\sigma/2,
\end{equation}
where
\begin{equation}
n=\sum_\sigma\langle c^\dag_{i\sigma}c^{}_{i\sigma}\rangle,\;
m=\sum_\sigma\sigma \langle c^\dag_{i\sigma}c^{}_{i\sigma}\rangle
\end{equation}
are the electron density and the site magnetization.
Both $z^2_\sigma$ and $\lambda_\sigma$ are essentially spin dependent, which allows the magnetic states formation to be studied.
However, in the slave-boson formalism, in contrast with HFA, the one-electron energy shift $\lambda_\sigma$ cannot be
expressed in terms of $n$ and $m$ only and should be specified separately, see Eq. (\ref{eq:lmb_eq}) below.

Then we can write the saddle-point (mean-field) equations:
\begin{eqnarray}
	\label{eq:n_eq}
	n &=& \frac1{N}\sum_{\kk s}f(E_s(\kk)),\\
	\label{eq:m_eq}
	m &=& \frac1{N}\sum_{\kk s}sf(E_s(\kk))\frac{e^+_\kk(z^2_\uparrow-z^2_\downarrow)+\lambda_\uparrow-\lambda_\downarrow}{\sqrt{D_\kk}},\\
	\label{eq:lmb_eq}	
	\lambda_\sigma &=&\nu\left[ \Phi_{\sigma}\left(\frac{p_{\bar\sigma}/e}{e^2+p_{\bar\sigma}^2} + \frac{d/p_{\sigma}}{p_{\sigma}^2+d^2}\right)+\Phi_{\bar\sigma}/(ep_\sigma)\right],
\end{eqnarray}
where we have introduced $\nu = ed-p_\uparrow p_\downarrow$, $f(E)$ = $(\exp((E-\mu)/T)+1)^{-1}$ is the Fermi function, $\mu$ is the chemical potential, $N$ is the lattice site number. We have introduced the quantity 
\begin{equation}
	\label{eq:Phi_def}
	 \Phi_{\sigma}\equiv\frac{ep_\sigma+p_{\bar\sigma}d}{(e^2+p_{\bar\sigma}^2)(p_{\sigma}^2+d^2)}\frac1{N}\sum_{\kk s}f(E_s(\kk))\frac{\partial E_s(\kk)}{\partial (z^2_{\sigma})},
\end{equation}	
which is an analogue of the Harris--Lange shift~\cite{Harris-Lange1967}, and the derivative $\partial E_s(\kk)/\partial (z^2_{\sigma})$ corresponds to the formal derivative of Eq. (\ref{eq:subband_spectrum}) with respect to $z^2_\sigma$.

The saddle-point values of bosonic variables should be obtained as a solution of the self--consistent equations:
\begin{eqnarray}
	\label{eq:n_func}
	2d^2+p_\uparrow^2+p_\downarrow^2&=&n,\\
	\label{eq:m_func}
	p_\uparrow^2-p_\downarrow^2&=&m,\\
	\label{eq:constraint}
	e^2+p_\uparrow^2+p_\downarrow^2+d^2&=&1,\\
	\label{eq:main_p}
	\nu\sum_\sigma \left(\frac1{ep_{\bar\sigma}}+\frac1{p_\sigma d}\right)\Phi_\sigma &=& U.
\end{eqnarray}

We consider now formal relation between Hartree-Fock and  slave-boson approximations. Note that the equations for HFA can be  represented in a form similar to SBA, since the former one treats the double occupancy as
\begin{equation}
d^2 = \langle d^\dag_i d_i\rangle = \langle n_\uparrow n_\downarrow\rangle \rightarrow  \langle n_\uparrow\rangle \langle n_\downarrow\rangle,
	\label{eq:HF_ansatz}
\end{equation}
which can be replaced in virtue of Eqs.~(\ref{eq:n_func}) and (\ref{eq:m_func}),
\begin{equation}
d^2 = (p^2_\uparrow+d^2)(p^2_\downarrow + d^2),
\end{equation}
which implies
\begin{equation}
d^2 (e^2) = \frac{1-p^2_\uparrow - p^2_\downarrow}2\pm\sqrt{\left(\frac{1-p^2_\uparrow-p^2_\downarrow}2\right)^2-p^2_\uparrow p^2_\uparrow},
	\label{eq:d_eq}
\end{equation}
where the signs are chosen to satisfy $d^2<e^2$ for $n<1$ and vice versa.
This replacement of Eq. (\ref{eq:main_p}) by Eq. (\ref{eq:d_eq}) allows to formulate HFA approximation in terms of bosonic quantities.
Now one can see that the HFA  differs from SBA because the former neglects the different nature of the singly
and doubly occupied states forming the electron spectrum.
The rough ansatz (\ref{eq:HF_ansatz}) does not include the bandwidth narrowing ($z^2_{\sigma}$ factor) caused by Pauli
principle, which is taken into account within SBA: 
Eq. (\ref{eq:d_eq}) implies that $\nu = ed-p_\uparrow p_\downarrow = 0$ and, from Eq.~(\ref{eq:z_def2}),  $z^2_\uparrow = z^2_\downarrow = 1$.
However, Eq. (\ref{eq:main_p}) means that $\nu=0$ implies $\lambda_\sigma=0\ne\lambda_\sigma^{\rm HF}$, which manifests the inconsistency of Eqs.~(\ref{eq:lmb_eq}) and (\ref{eq:HF_ansatz}) within HFA.
A concrete example for the concentration dependences of bosonic fields defined above within HFA  and SBA is presented in Sect. \ref{sect:analysis}.

The saddle-point expression for the spiral state thermodynamic potential (per site) $\Omega(\mu,\mathbf{Q})$ has
the form

\begin{equation}
	\label{eq:Omega_sdl_pt}
	\Omega(\mu,{\bf Q})=Ud^2-\sum_\sigma\lambda_\sigma(p_\sigma^2+d^2)+\Omega_{\rm f}(\mu,z^2_\sigma,\lambda_\sigma),
\end{equation}
where $\Omega_f(\mu,z^2_\sigma,\lambda_\sigma)$ is the standard potential for the non--interacting electron system in the
field $\lambda_\sigma$ with a narrowed (by the factor $z^2_\sigma$) spectrum
\begin{equation}
	\label{eq:Omega_f_definition}
	\Omega_{\mu,\rm f}(z^2_\sigma,\lambda_\sigma) \equiv -(T/N)\sum_{\kk s}\log\left(1+\exp(-(E_s(\kk)-\mu)/T)\right) .
\end{equation}

The resulting wave vector is determined by the minimization of $\Omega$ over various spiral states (the spiral wave vector
$\bf{Q}$) at fixed $\mu$ in the limit $T\rightarrow0$. The minimization of $\Omega(\mathbf{Q})$ with respect to $\bf Q$ was performed
numerically with $\textbf{Q}$
running in high-symmetry directions of the Brillouin zone (we choose the lattice spacing as unity): the $(Q,Q)$, $(Q,\pi)$ and $(0,Q)$ directions were considered
for the square lattice; the $(Q,Q,Q)$, $(Q,Q,\pi)$, $(Q,\pi,\pi)$, $(0,0,Q)$, $(0,Q,Q)$ and $(0,Q,\pi)$ directions were taken
into account for cubic lattices (for bcc and fcc lattices $\pi$ is replaced by 2$\pi$ due to halving the lattice
spacing).
The step of changing $Q$ was $0.02\pi$ for the square lattice and $0.05\pi$ for cubic lattices. The number of
$\kk$-points in the Brillouin zone during integration was 400 for the square lattice and 70 for cubic lattices. The accuracy
settings were adjusted, if necessary. Since $\Omega$  in the ground state actually depends on the chemical potential
$\mu$ as a parameter, we can determine the dependence of the magnetic structure on $\mu$  by a procedure taking into
account the possibility of the phase separation~\cite{Igoshev10}, avoiding the tedious application of Maxwell rule in the
negative compressibility regions.

To illustrate the non--uniform state formation we present an example of the density dependence of the chemical potential, calculated by commonly used method and the present method (see also Ref.~\onlinecite{Igoshev10}) in the vicinity of half--filling. One can see that the former method yields a thermodynamical unstable dependence $\mu(n)$ which should be corrected by additional applying the Maxwell rule. 
\begin{figure}[h!]
\center
\includegraphics[width=0.5\textwidth]{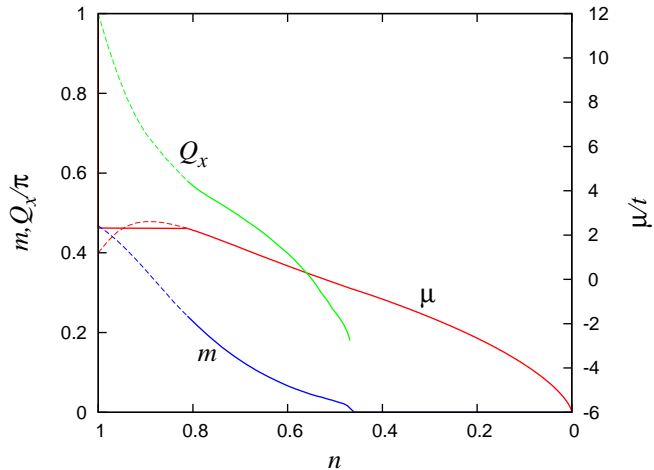}
\caption{
	(Color online) The dependence of the chemical potential $\mu$, magnetization $m$ and $Q_x$ vs electonic density for simple cubic lattice, calculated within SBA, $t'=0$, $U = 14t, \mathbf{Q}=(Q_x,\pi,\pi)$. Solid line presents the result with account of PS, dashed line is the standard result of Ref.~\onlinecite{Fresard92}. The choice of the wave vector form provides  the minimum of the total energy (see Fig. \ref{fig:cubic_t'=0} below). 
}
\label{fig:mu_vs_n}
\end{figure}

\section{Fermi surface nesting and van Hove singularities}
\label{sec:nesting}

Before proceeding to the results obtained we need to mention the main physical factors that are important for the incommensurate magnetic
states formation starting from paramagnetic phase. There are two well known mechanisms leading to stabilization of incommensurate states: (i) Fermi surface nesting and (ii) van Hove scenario.

(i) Fermi surface nesting is the presence of large segments of a Fermi surface that can be connected to other segments
via a nesting vector $\mathbf{Q}$ in such a way that occupied states move to the empty states region and vice versa.
When the nesting condition $\epsilon_{\mathbf{k}}=-\epsilon_{\mathbf{k+Q}}$ is realized for many $\kk$-points the
magnetic susceptibility function
\begin{equation}
\label{eq:susceptibility}
\chi_0(\mathbf{Q})=\frac1{N}\sum_{\kk}\frac{f(\epsilon_\kk)-f(\epsilon_{\kk+\mathbf{Q}})}{\epsilon_{\kk+\mathbf{Q}}
-\epsilon_\kk}
\end{equation}	
is enhanced or has a singularity leading to stabilization of the incommensurate magnetic state with wave vector $\mathbf{Q}$ \cite{Lomer}.
One can mention that the nesting condition means $e_\kk^+=0$ in our formalism (see (\ref{eq:esa})).  This results
in cancellation of the symmetric terms in spectra $E(\kk)$ (\ref{eq:subband_spectrum}), and hence the
$z$-dependence of $E(\kk)$ becomes appreciably suppressed. Thus we come to an important conclusion that the Fermi
surface
nesting makes the system to some extent insensitive to the correlation effects. In the FM case
$\mathbf{Q}=0$ and the nesting condition reduces to $\epsilon_{\mathbf{k}}=0$. This means that the larger is the density of
states at a given $\mu$, the weaker is the influence of electron correlations on the FM state.

(ii) Another mechanism leading to the incommensurate magnetic structure formation was proposed in Ref.~\onlinecite{Rice75}. The
authors have shown that the presence of saddle points in the electron spectrum results in the divergence of susceptibility
(\ref{eq:susceptibility}) if $\mathbf{Q}$ is equal to a vector connecting the saddle points to each other. The result was
obtained for the 2D case but one can expect that it holds true for the 3D case as well with a possible change of the
divergence into a weak singularity. This mechanism demonstrates the importance of van Hove singularities (given by
the saddle points) for magnetism.

Thus, when analyzing the magnetic phase diagrams, special attention should be paid to the density of states, nesting
properties of the Fermi surface, and van Hove singularities.

\section{Results: square lattice}
\label{sec:results_2D_lattices}

\subsection{$t'=0$}
\begin{figure}[h]
\center
\includegraphics[width=0.5\textwidth]{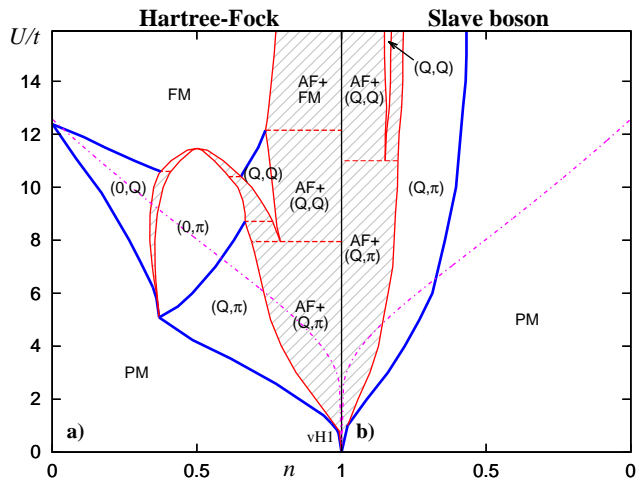}
\caption{
	(Color online)
	Ground state magnetic phase diagram of the Hubbard model for the square lattice with $t'=0$ at $n<1$ within (a) HFA~\cite{Igoshev10} and (b) SBA.
	The spiral phases are denoted according to the form of their wave vector.
	Filling shows the phase separation regions.
	Bold (blue) lines denote the second-order phase transitions.
	Solid (red) lines correspond to the boundaries between the regions of the homogeneous phase and phase
separation.
	Dashed (red) lines separate the PS regions corresponding to different phase pairs.
	The second-order phase transition boundary produced by the Stoner criterion $U_{\rm c} = 1/\rho(E_{\rm
F})$ (where $\rho(E_{\rm F})$ is the density of states at the Fermi level in the paramagnetic phase), separating the ferromagnetic and paramagnetic regions, is shown by dash--dotted (violet)
line.
	$\mathbf{Q}_{\rm AFM} = (\pi,\pi)$.
}
\label{fig:square_t'=0}
\end{figure}

We calculated the phase diagrams of the ground state at different $t'/t$ values within the Hartree--Fock approximation
in~Ref.~\onlinecite{Igoshev10}; the results for $t' = 0$ are presented in Fig. \ref{fig:square_t'=0}a.
An analogous phase diagram obtained within SBA is shown in Fig.~\ref{fig:square_t'=0}b~\cite{Igoshev2013} (we restrict ourselves to the case $0<n\le1$ due to the particle--hole symmetry, see justification below).
Since SBA takes into account the impact of many-electron states onto the one-electron spectrum on the static level, we can treat these SBA corrections as correlation effects beyond HFA. 
Note that for both these approaches the phase transitions between  different magnetic states are generally
first-order transitions, which leads to considerable regions of phase separation. 
This also applies to the transition from AFM to spiral state at arbitrary small doping, which is of the first order due to that the insulator finite-gap state goes abruptly to the metallic phase (see also \cite{Dzierzawa92}).
The account of PS substantially distinguishes the phase diagram obtained within SBA from that presented in Ref.~\onlinecite{Fresard92}, where PS was disregarded. 
The separation regions between the AFM phase and the spiral magnetic states (parallel, $\mathbf{Q}=(Q,\pi)$, and
diagonal, $\mathbf{Q}=(Q,Q)$) are rather wide; the
corresponding regions of pure spiral states are narrowed. In particular, this refers to the diagonal phase, the existence of which becomes possible only at $U > 11t$.
The phase transition between the PM and spiral magnetic states is of the second order. 
The magnitude of {\it number} $Q$ in the regions of spiral states is variable but qualitatively it behaves as shown in Fig.~\ref{fig:mu_vs_n}: the larger is the doping, the smaller is the value of $Q$. This behavior of $Q$ is typical for all the lattices considered.

One can see that the electron correlations lead to a noticeable suppression of the magnetically ordered states in comparison
with HFA: the corresponding concentration intervals in the phase diagram decrease strongly, and the variety of the
spiral states disappears.
Besides, in the slave-boson approximation there occurs a wide region of PM state.
The FM state covering a considerable part of the diagram within HFA is shifted to the region of large values
$U\gtrsim60t$.
This is a manifestation of the fact that SBA yields a convenient scheme of interpolation between the small $U$ limit
(coinciding with HFA) and the large $U$ limit ($d=0$ as the limit consequence of Eq.~(\ref{eq:main_p})) in the case $n<1$;
whereas HFA yields only saturated ferromagnetism at large $U$ for finite doping.
The suppression of FM state reproduces the result obtained in Ref.~\onlinecite{Fresard92} and is in  good agreement
with the variational study of the stability of the saturated FM phase~\cite{Linden91}.
The region of separation of the AFM and spiral phases is narrowed by about one half (with respect to the variable $n$).

According to our calculations, even an unrestricted increase of $U$ does not make the magnetically ordered states stable
far from half-filling: at $U\rightarrow\infty$, there are no spiral magnetic solutions of the slave-boson method equations at
$|1-n| > 0.63$.
At the same time, the saturated FM solution becomes more favorable than the spiral ones only at $|1-n| < 0.3$.
This improves the result proposed in Ref.~\onlinecite{Kotliar86}, where spiral phases were not considered; similar phase diagram was obtained in Ref.~\onlinecite{Fresard92} but PS was disregarded. 
Thus, at large $U$ values far from half-filling the spiral magnetic state  replaces the saturated FM one.
In contrast to Ref.~\onlinecite{Irkhin04}, in our approach there exist unsaturated FM solutions, but they are always energetically
unfavorable in comparison with the saturated FM or spiral magnetic states.

In order to compare our results with the Stoner criterion and analyze the van Hove singularities, the inverse density of states (DOS) at the Fermi level is depicted in Fig. \ref{fig:square_t'=0} and all diagrams below. 
There is a saddle point in the electron spectrum of the square lattice  $\mathbf{k}=(0,\pi)$ which produces a
divergent van Hove singularity of DOS. For $t'=0$ this singularity is at the band center (we introduce the notation
for the density corresponding to a van Hove singularity in the paramagnetic spectrum $n_{\rm
vH1} = 1$) and coincides with the case of perfect nesting. Both these peculiarities determine the weakness of correlation effects in the vicinity of half-filling, cf.~Fig.~\ref{fig:mu_vs_n}, which is reflected in qualitative accordance of the HFA and SBA results close
to $n_{\rm
vH1}$.

\subsection{$t'=0.2t$}

\begin{figure}[h]
\includegraphics[width=0.49\textwidth]{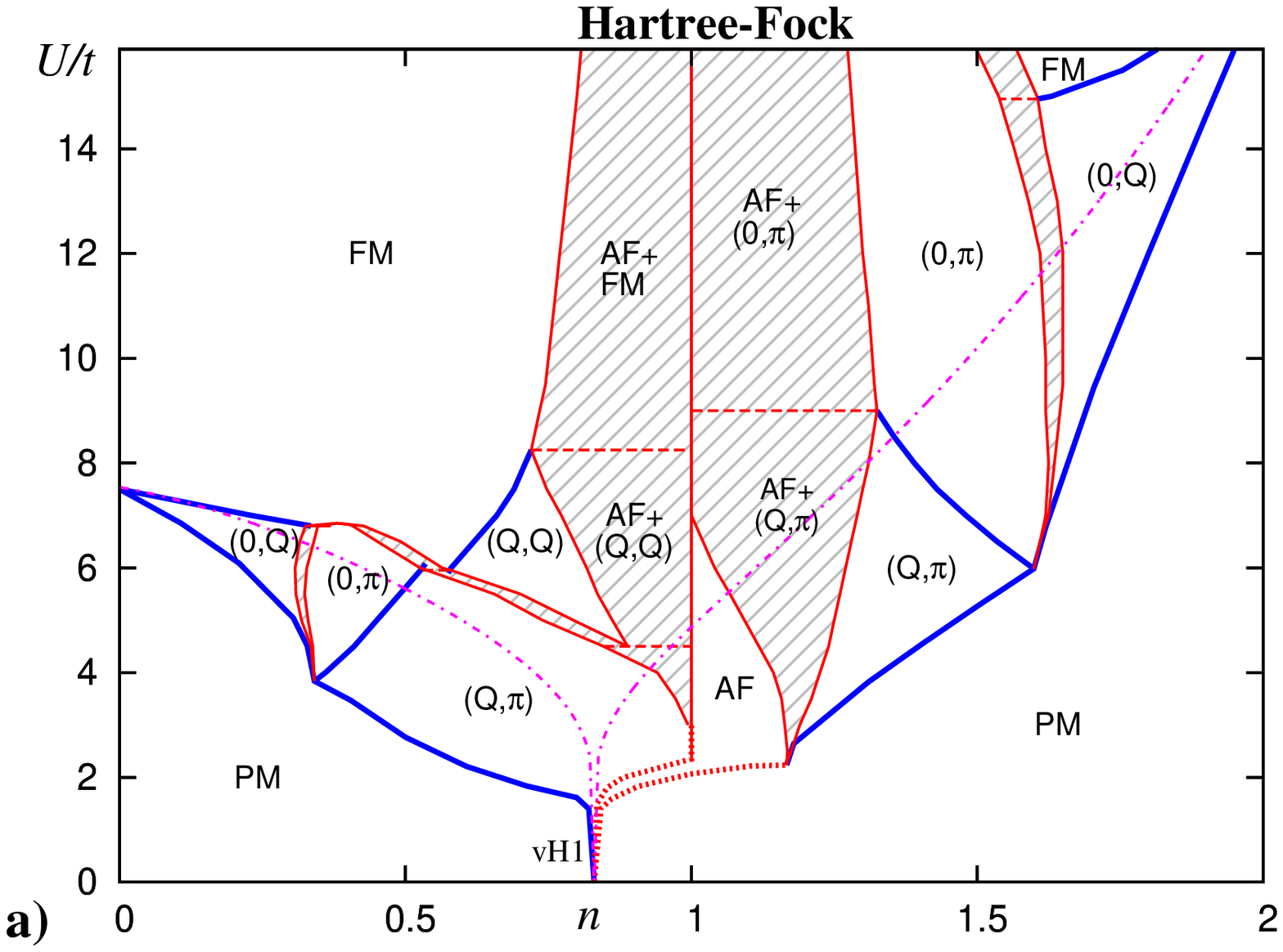}
\includegraphics[width=0.49\textwidth]{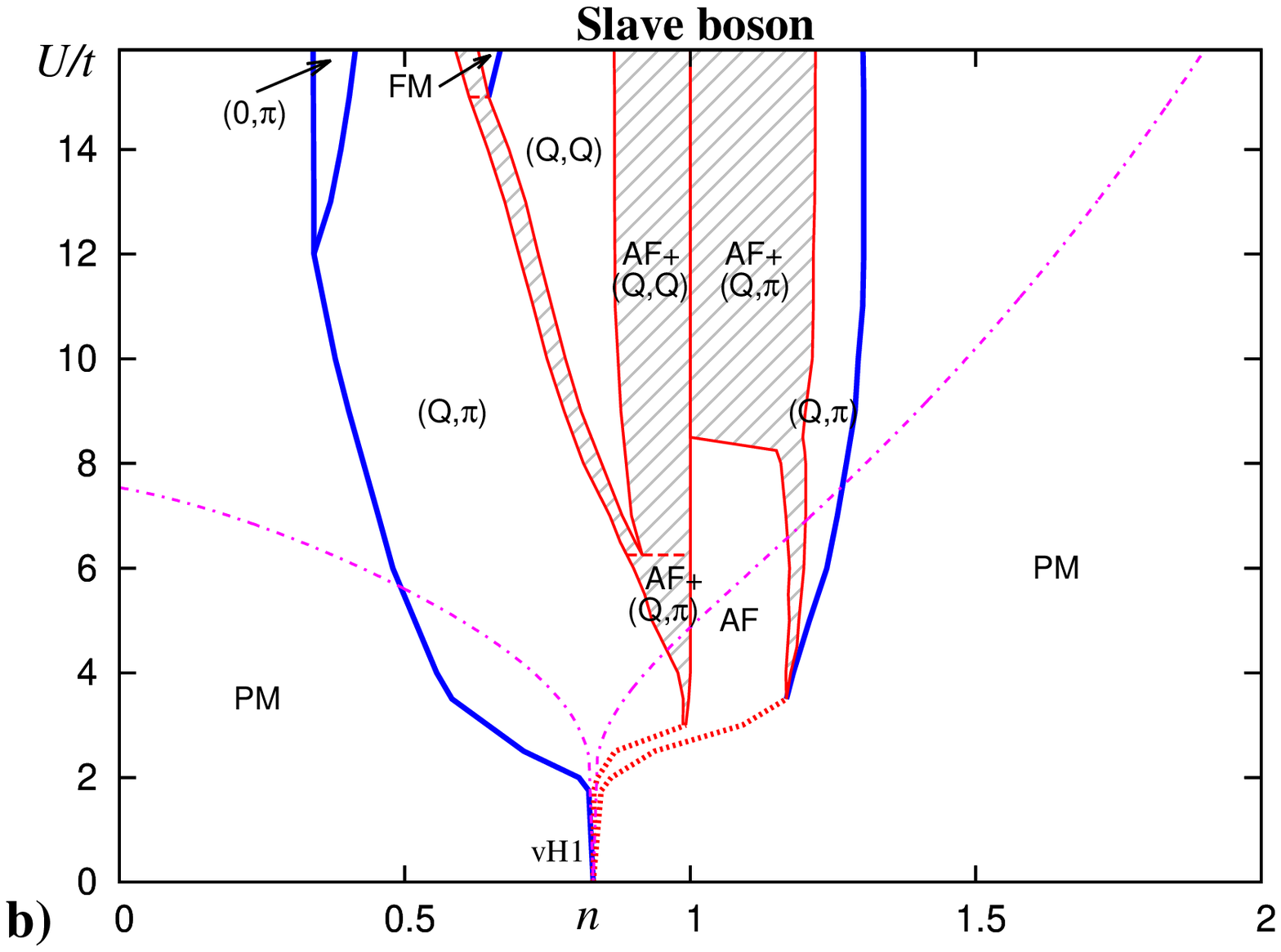}
\caption{
	(Color online)
	Ground state magnetic phase diagram of the Hubbard model for the square lattice with $t'=0.2t$ within (a) HFA~\cite{Igoshev10}; (b) SBA.
	Bold dotted (red) lines denote the first order phase transitions if the phase separation region is narrow.
	The notations are analogous to those in Fig.~\ref{fig:square_t'=0}.
}
\label{fig:square_t'=0.2}
\end{figure}

Up to the end of this Section we present our results for the square lattice for the nonzero value of $t'$ at which the
particle--hole symmetry breaks down: of great interest is the difference between the properties of the hole-doped
($n<1$) and electron-doped ($n>1$) systems \cite{Igoshev10}.
The consideration of the negative ratio $t'/t$ can be reduced, as a consequence of lattice bipartiteness, to that of the positive one by the particle--hole transformation $c^{}_{i\sigma}\rightarrow(-1)^ic^\dag_{i\sigma}, t'\rightarrow -t', n\rightarrow2-n$, where the site numbering provides different parity for neighboring sites.
We present the phase diagram for $t'=0.2t$ within HFA (Fig. \ref{fig:square_t'=0.2}a) and SBA (Fig. \ref{fig:square_t'=0.2}b).
In comparison with the case of $t'=0$, for $n<1$ the diagonal phase shifts to the region of much smaller $U/t$ values and the parallel phase becomes more extended over the concentration parameter.
The physical cause is the shift of the density value $\nVH$ (for $t'=0.2t$ $n_{\rm vH1}\approx0.83$) from the
half-filling ($n = 1$):
at $t'\ne0$, $\nVH$ and the half-filling  determine two different trends to the corresponding magnetic orderings.
We see that in the hole-doped half of the phase diagram the correlation effects lead only to the quantitative
renormalization of the phase boundaries.

In the case of $n > 1$, the correlation effects are more significant: all homogeneous spiral states disappear, except for a narrow region of the parallel phase.
Such a suppression of magnetism can be explained by the fact that in this case all peculiarities (imperfect nesting)  come
from the point  $n = 1$.
Far from half-filling, no spiral magnetism is possible and the saturated ferromagnetism is energetically unfavorable in
comparison with the PM phase at any large $U$ value, when correlations are taken into account.
Thus, the particle--hole asymmetry is enhanced considerably as compared with HFA.

These results can be used for qualitatively explaining  the magnetic properties of layered high-temperature
superconducting perovskites, for which the fitting of the angular resolved photoemission spectroscopy spectra to those
of the bare model yields $t'\sim 0.2t$, see \cite{Igoshev10,Igoshev2013} and references therein. 
The results are in agreement with the experimental data on the magnetic structure of the hole-doped
compound La$_{2-p}$Sr$_p$CuO$_4$ which has a close value of the $t'/t$
parameter~\cite{Tanamoto93,Matsuda02,Fujita02,Yamada98}.
At the same time, for the high-temperature superconducting compound Nd$_{2-x}$Ce$_x$CuO$_4$, in which the charge
carriers are electrons, the homogeneous commensurate AFM ordering is stable up to $x =
0.14$~\cite{Takagi89} in agreement with our results for $n > 1$ (see Fig. \ref{fig:square_t'=0.2}).

\section{Results: 3D lattices}
\label{sec:results_3D_lattices}
Now we present  the results for three-dimensional lattices: simple cubic (sc), body-centered cubic (bcc), and
face-centered cubic (fcc) lattices. 
We restrict ourselves to the collinear and spiral structures with magnetic cell being not increased, so that.,e.g., type--III antiferromagnetic collinear structure with a doubled cell \cite{Goodenough_book} is beyond our consideration; such structures will be treated elsewhere. 
\subsection{Simple cubic lattice}
\label{sec:sc}
\subsubsection{$t'=0$}
\begin{figure}[h!]
\center
\includegraphics[width=0.5\textwidth]{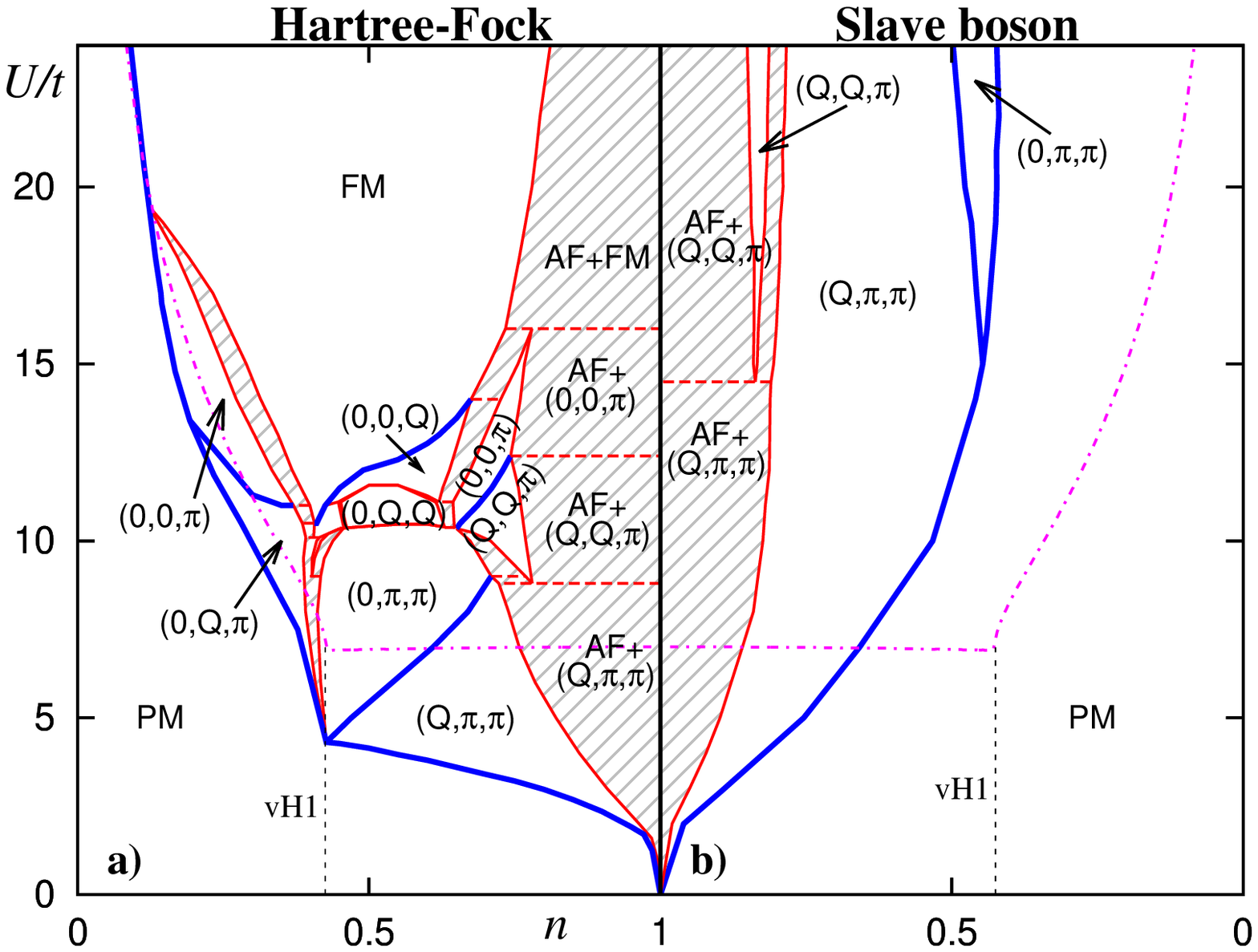}
\caption{
	(Color online) The same as in Fig. \ref{fig:square_t'=0} for the simple cubic lattice Hubbard model with
$t'=0$.
	$\mathbf{Q}_{\rm AFM} = (\pi,\pi,\pi)$.
}
\label{fig:cubic_t'=0}
\end{figure}

The physical picture is very similar to that for the square lattice (see Fig.
\ref{fig:cubic_t'=0} for a comparison of the results of HFA (a) and SBA (b)
approaches): the density value $n=1$ corresponding to the perfect AFM
nesting peculiarity (but not to a van Hove singularity) retains its crucial role.
The spiral magnetic phase with $\textbf{Q} = (Q,\pi)$ of the square lattice is
replaced by a spiral with ${\bf Q}=(Q,\pi,\pi)$.
The electron spectrum of the sc lattice has two saddle $\kk$-points: the X and M points of the Brillouin zone which give two van Hove
singularities of DOS at the electron concentrations $n_{\rm vH1}\approx0.425$ and $n_{\rm vH2}\approx1.575$, respectively. The vector connecting two X-points with coordinates $(0,0,\pi)$ and $(0,-\pi,0)$ determines the tendency to $\mathbf{Q}=(0,\pi,\pi)$ state by the van Hove scenario described in Section \ref{sec:nesting}.
In contrast to the 2D case, the van Hove singularities are bounded and do not much affect the SBA diagram --- the correlation effects
strongly suppress magnetism at $n$ being far from half-filling, including the regions close to $n_{\rm vH}$'s.

\subsubsection{$t'=0.3t$}
\begin{figure}[h!]
\includegraphics[width=0.49\textwidth]{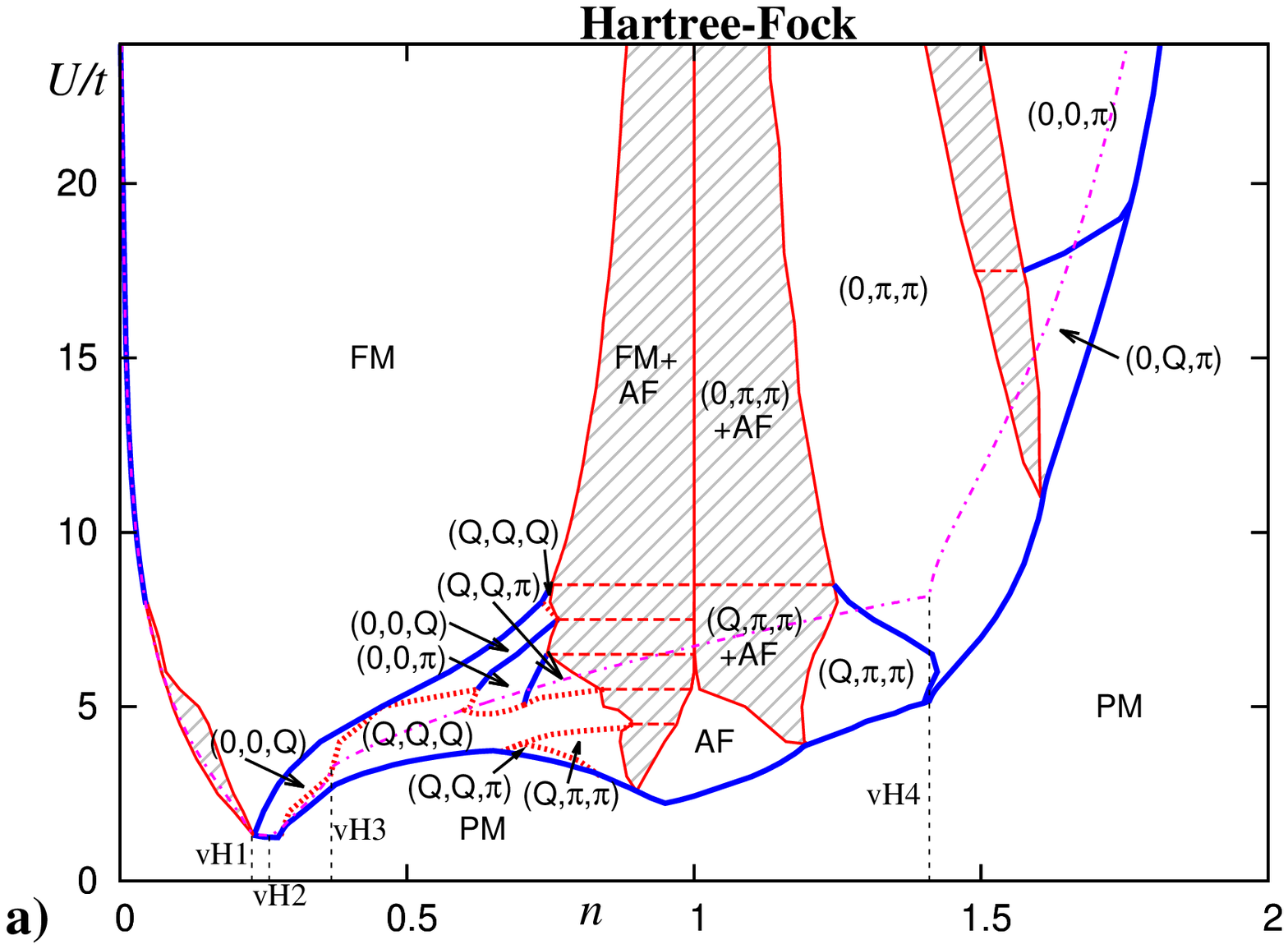}
\includegraphics[width=0.49\textwidth]{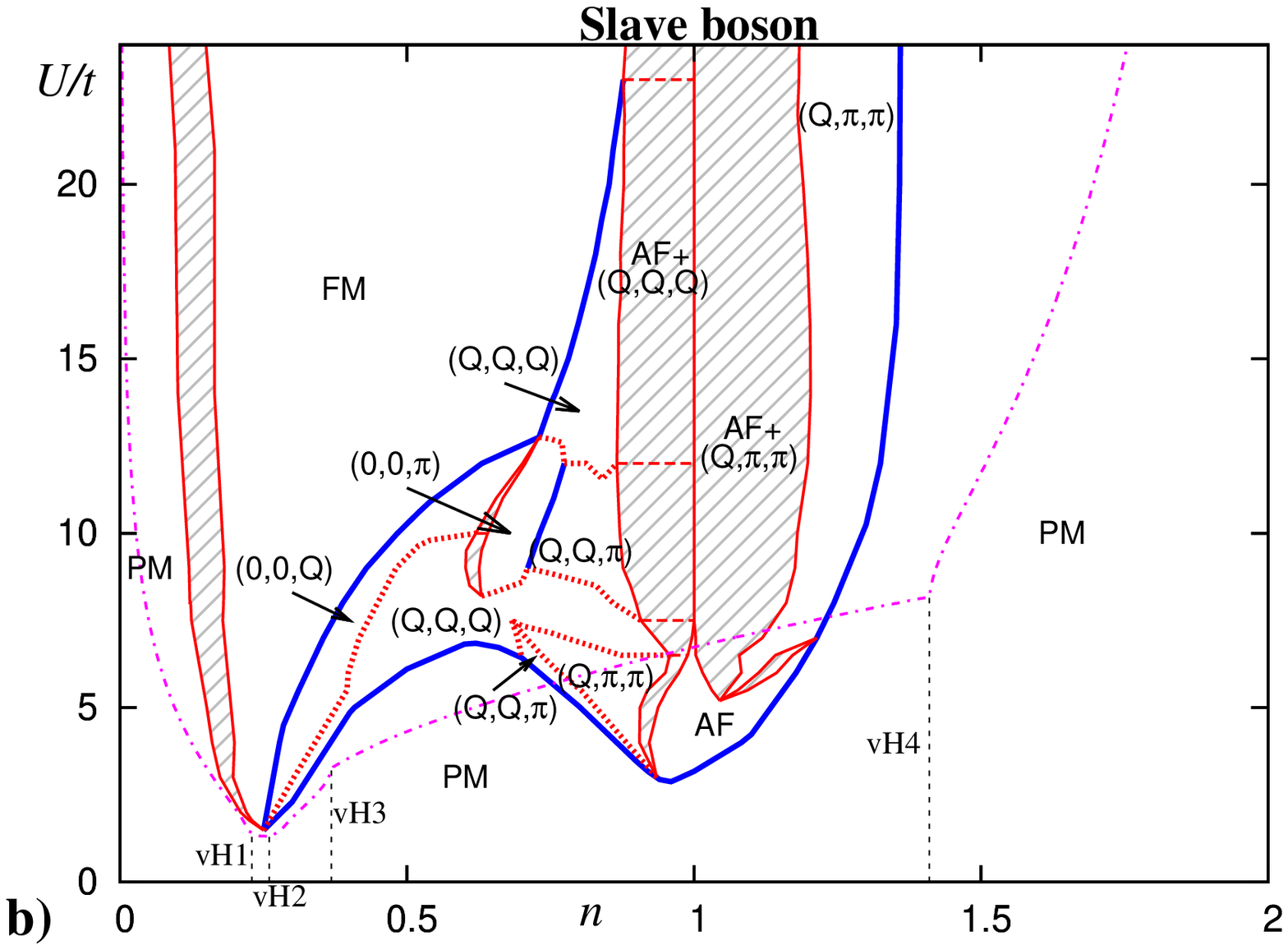}
\caption{
  	(Color online) The same as in Fig. \ref{fig:square_t'=0.2} for the simple cubic lattice Hubbard model with
$t'=0.3t$ (a) HFA results; (b) SBA results.
  $\mathbf{Q}_{\rm AFM} = (\pi,\pi,\pi)$.
}
\label{fig:cubic_t'=0.3}
\end{figure}

In Figs. \ref{fig:cubic_t'=0.3}a and \ref{fig:cubic_t'=0.3}b are presented the respective HFA and SBA results for the
case of positive non-zero $t'/t$, when
these tendencies come out in different parameter regions on the phase diagram
(we consider the case of $t'=0.3t$, the consideration of the negative ratio $t'/t$
is omitted for the same reason as in Section \ref{sec:results_2D_lattices}).

\begin{figure}[h!]
\center
\includegraphics[width=0.5\textwidth]{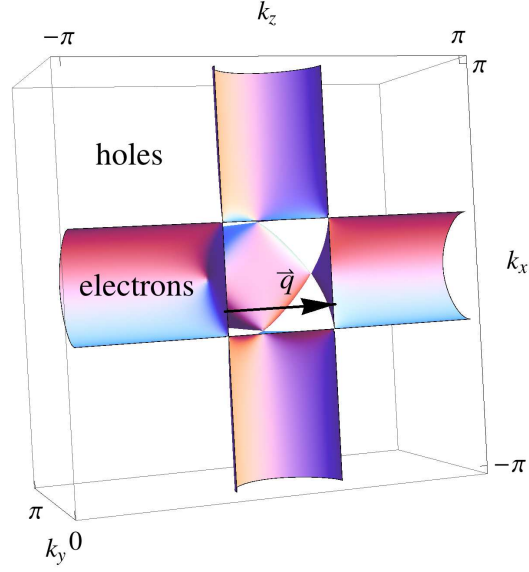}
	\caption{(Color online) Fermi surface section in a half of Brillouin zone for sc lattice, $t'=0.3t$, $n=n_{\rm
vH1}$. $\mathbf{q}$ is the nesting vector, see text.
}
\label{fig:nesting}
\end{figure}

This
case is characterized by four van Hove singularities (all being bounded): $n_{\rm vH1}\approx0.23$
($\kk=(0,\alpha,\alpha)$, $\cos\alpha=t/2t'-1$); $n_{\rm vH2}\approx0.26$
($\kk=(\beta,\beta,\beta)$, $\cos\beta=t/4t'$); $n_{\rm
vH3}\approx0.368$ ($\Gamma$ point); $n_{\rm vH4}\approx1.41$ (M point).
The saddle points which give the 'vH1' and 'vH2' singularities do not correspond to special high-symmetry points of the
Brillouin zone and hence the van Hove scenario described in Sec. \ref{sec:nesting} favors the formation of incommensurate vertical
($\mathbf{Q}=(0,0,Q)$) and diagonal ($\mathbf{Q}=(Q,Q,Q)$) states.   The reduced symmetry of the saddle points
leads to the removal of their degeneracy and, as a result, to an increase of the DOS. The large DOS value stabilizes the FM phase for a
rather small $U/t\approx 1.5$ and protects it from suppression by the electron correlations (see the SBA diagram). Also the spiral
states turn out to be stable to correlation effects, which can be explained by the Fermi surface nesting
formation. The Fermi surface section for the case where the paramagnetic state density $n = n_{\rm vH1}$ is depicted in Fig. \ref{fig:nesting}. It consists of six
cylindrical figures with regular squares at their bases. These squares are connected by the nesting vector
$\mathbf{q}=(0,0,q), q=2\arccos(t/2t'-1)\approx0.267\pi$. In accordance with the nesting concept this leads to magnetic
instability with respect to the formation of incommensurate structure with $\mathbf{q}=\mathbf{Q}$. A comparison with the calculated phase diagram (Fig.
\ref{fig:cubic_t'=0.3}) shows an agreement in the $Q$ value with the $(0,0,Q)$ phase found near $n_{\rm vH1}$.

As well as in the 2D case, we found a strong  hole--particle asymmetry: for $n<1$ a variety of magnetic phases appear
at not large $U/t$.
For instance, at a realistic value $U=4t$ we found the AFM phase in the vicinity of
half-filling. As the density lowers, it is replaced by a phase with $\mathbf{Q}=(Q,\pi,\pi)$ and
a small fracture of $(Q,Q,\pi)$ phase; on further lowering of the density a wide non-magnetic region is observed, and at
low
density the system enters into the region of competition between the FM, $(Q,Q,Q)$ and $(0,0,Q)$ phases.

\begin{figure}[h!]
\includegraphics[width=0.49\textwidth]{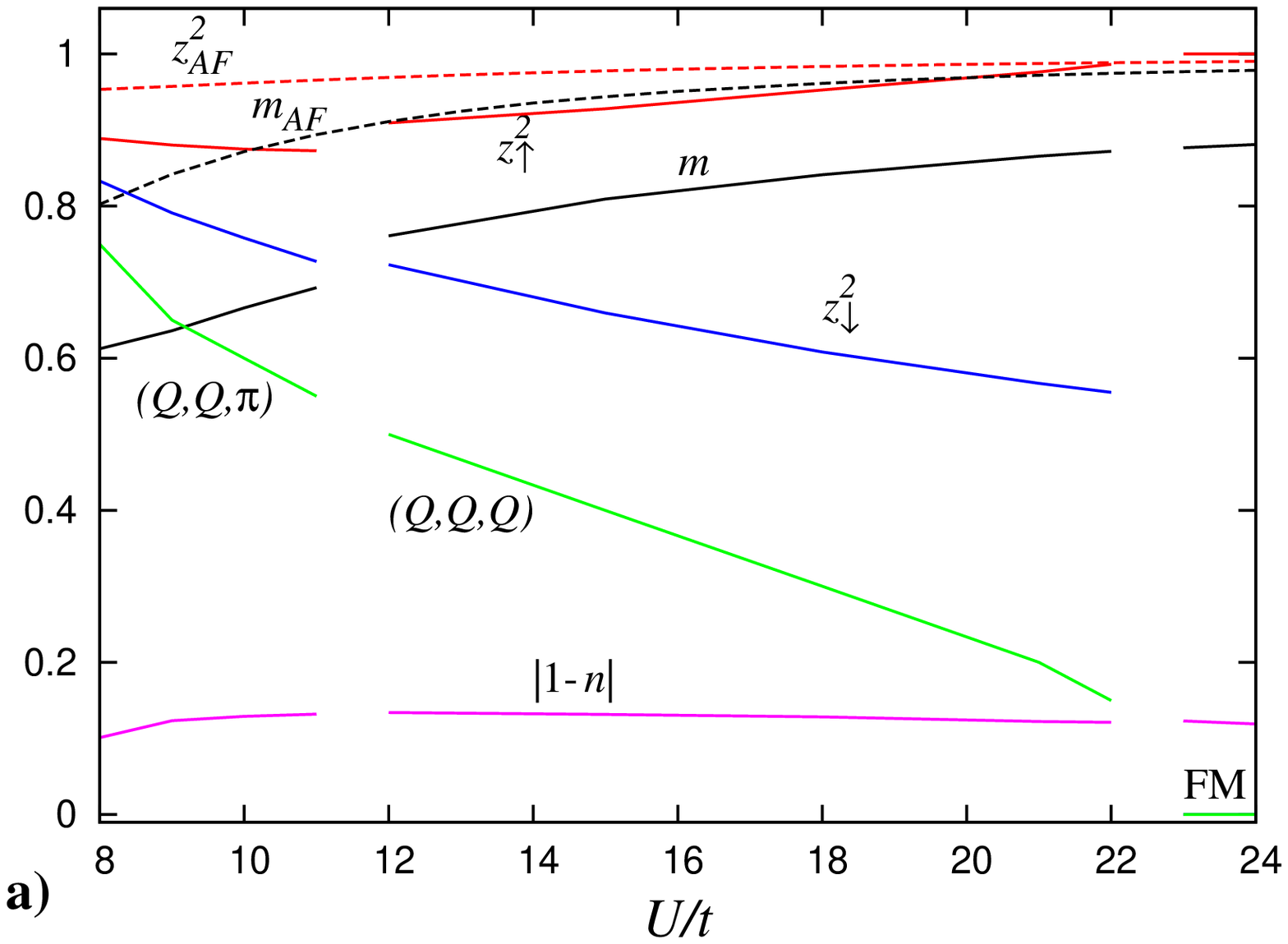}
\includegraphics[width=0.49\textwidth]{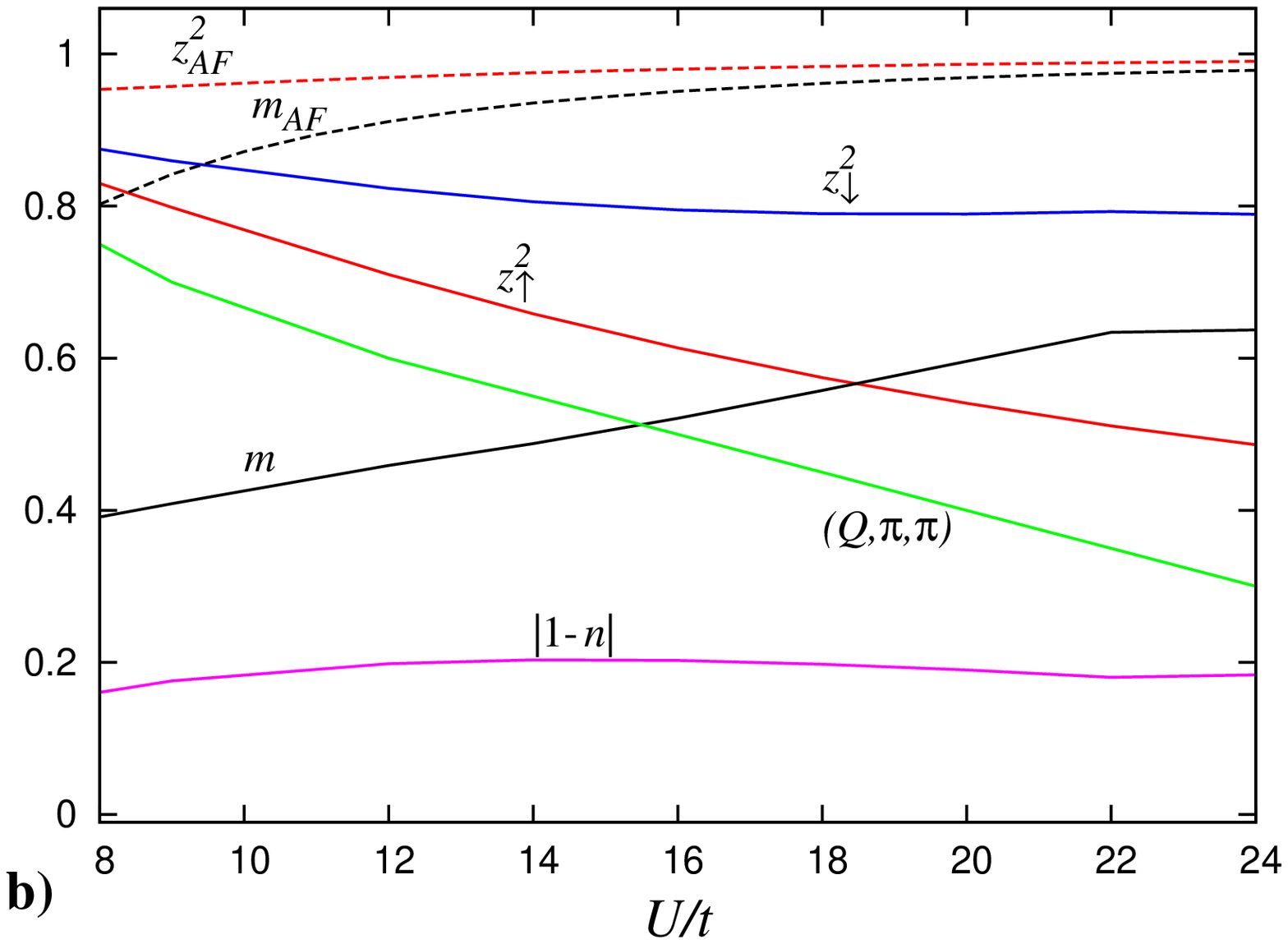}
\caption{
        (Color online) Ground state  characteristics, $z^2_\sigma$, $m$, 
$Q_x/\pi$ and $n$, of the phases participating in PS in the vicinity of 
half--filling for the sc lattice at $t'=0.3t$ as a function of $U$: (a) 
$n\le1$ and (b) $n\ge1$.
        Solid lines in (a) and (b) correspond to left and right boundary line of PS region, respectively; 
dashed lines present $m_{\rm AF} = m$ and $z^2_{\rm AF} = z^2_{\uparrow} = z^2_{\downarrow}$ for the AFM case at $n=1$. 
Line breaks in (a) originate from the first order transitions, being connected with the jump of the spiral state wave vector $\mathbf{Q}$.
The magnetization jump $m_{\rm AF}-m$ is given by the difference of dashed and solid lines.
}
\label{fig:PS_properties}
\end{figure}

To illustrate the properties of phase separation into N\'eel antiferromagnetic at $n=1$ and spiral (ferromagnetic) phase ($n \ne 1$) we present their $U$ dependence ($z^2_\sigma$, $m$, $\mathbf{Q}$ and $n$) along the boundaries of PS region (see Fig.~\ref{fig:cubic_t'=0.3}) for both these phases participating in PS within SBA in Fig.~\ref{fig:PS_properties}. 
As $U$ increases both $m$ and $z^2_{\uparrow}$ ($z^2_{\downarrow}$) for $n<1$ ($n>1$) tends to 1, on the other hand, $z^2_{\downarrow}$ ($z^2_{\uparrow}$) for $n<1$ ($n>1$) is strongly reduced.
The density $n$ at the boundary ($n\ne 1$) appears to be only weakly $U$--dependent. 
These tendencies are much stronger for the AFM phase at $n=1$, which implies that correlation effects are much more pronounced at $n\ne 1$ than in half--filled case $n=1$ where $z_\sigma$ are close to unity. 
The magnetization jump between these two phases turns out to be almost $U$--independent, but considerably different for $n<1$ ($|\Delta m|\simeq 0.2$) and $n>1$ ($|\Delta m|\simeq 0.4$). 

\subsection{Body-centered cubic lattice}
\label{sec:bcc}
\subsubsection{$t'=0$}
\begin{figure}[h!]
\center
\includegraphics[width=0.5\textwidth]{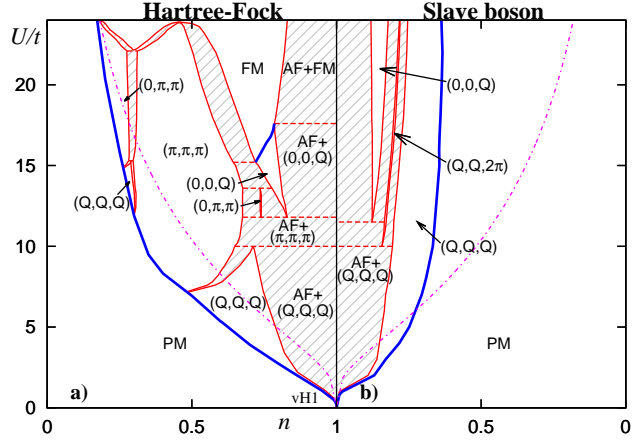}
\caption{
	(Color online)
	The same as in Fig. \ref{fig:square_t'=0} for the bcc lattice Hubbard model with $t'=0$,
	$\mathbf{Q}_{\rm AFM} = (0,0,2\pi)$.
}
\label{fig:bcc_t'=0}
\end{figure}

This lattice is characterized by the density of states with a strong $\log^2$-singularity at the band center
for $t'=0$, which corresponds to a whole line of van Hove singularities in the Brillouin zone ($\kk=(Q,\pi,\pi)$,
$0<Q<\pi$, D direction).
In Fig. \ref{fig:bcc_t'=0} we present the results for the bcc lattice Hubbard model for both HFA
(a) and SBA (b) at $0<n<1$ (the particle--hole symmetry holds).
The results are somewhat similar to those for the square lattice (logarithmic singularity).
One can see that the correlation effects strongly but only quantitatively renormalize the phase boundary lines in a
rather wide vicinity of half-filling, whereas well away from half-filling they fully destroy magnetic ordering.
The van Hove singularity corresponding to $n = \nVH = 1$, results in this case in especially wide regions of PS and
magnetic phases. This is a manifestation of the particularly strong bcc lattice van Hove singularity that merges with
the half-filling peculiarity related to the perfect nesting. The diagonal magnetic order ($\mathbf{Q}=(Q,Q,Q)$)
dominates, and at moderate  $U/t$ this is the only existing order.
In general the magnetic phase distribution is similar to the case of square lattice (see Fig. \ref{fig:square_t'=0}).

\subsubsection{$t'=0.3t$}
\begin{figure}[h!]
\includegraphics[width=0.49\textwidth]{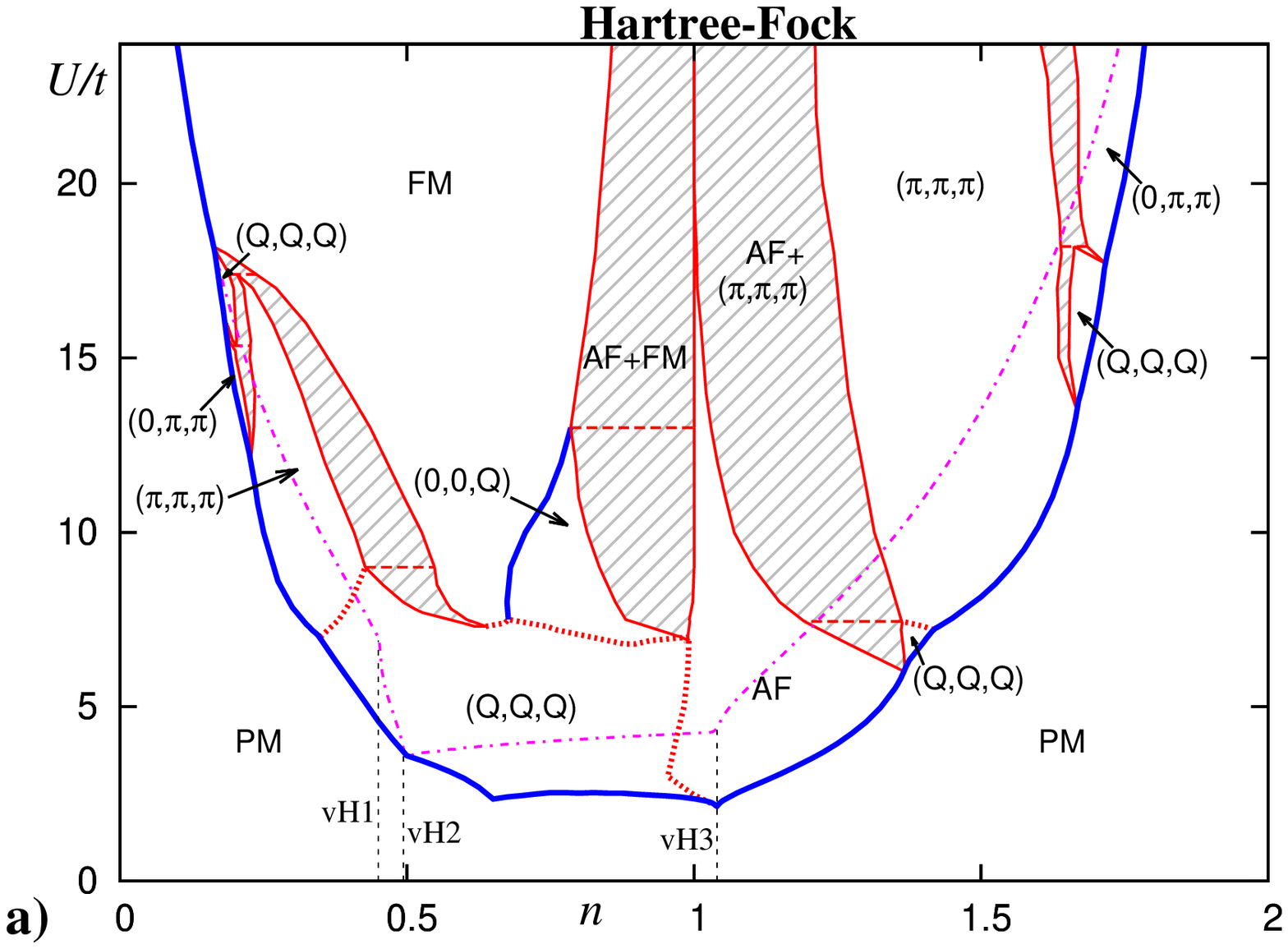}
\includegraphics[width=0.49\textwidth]{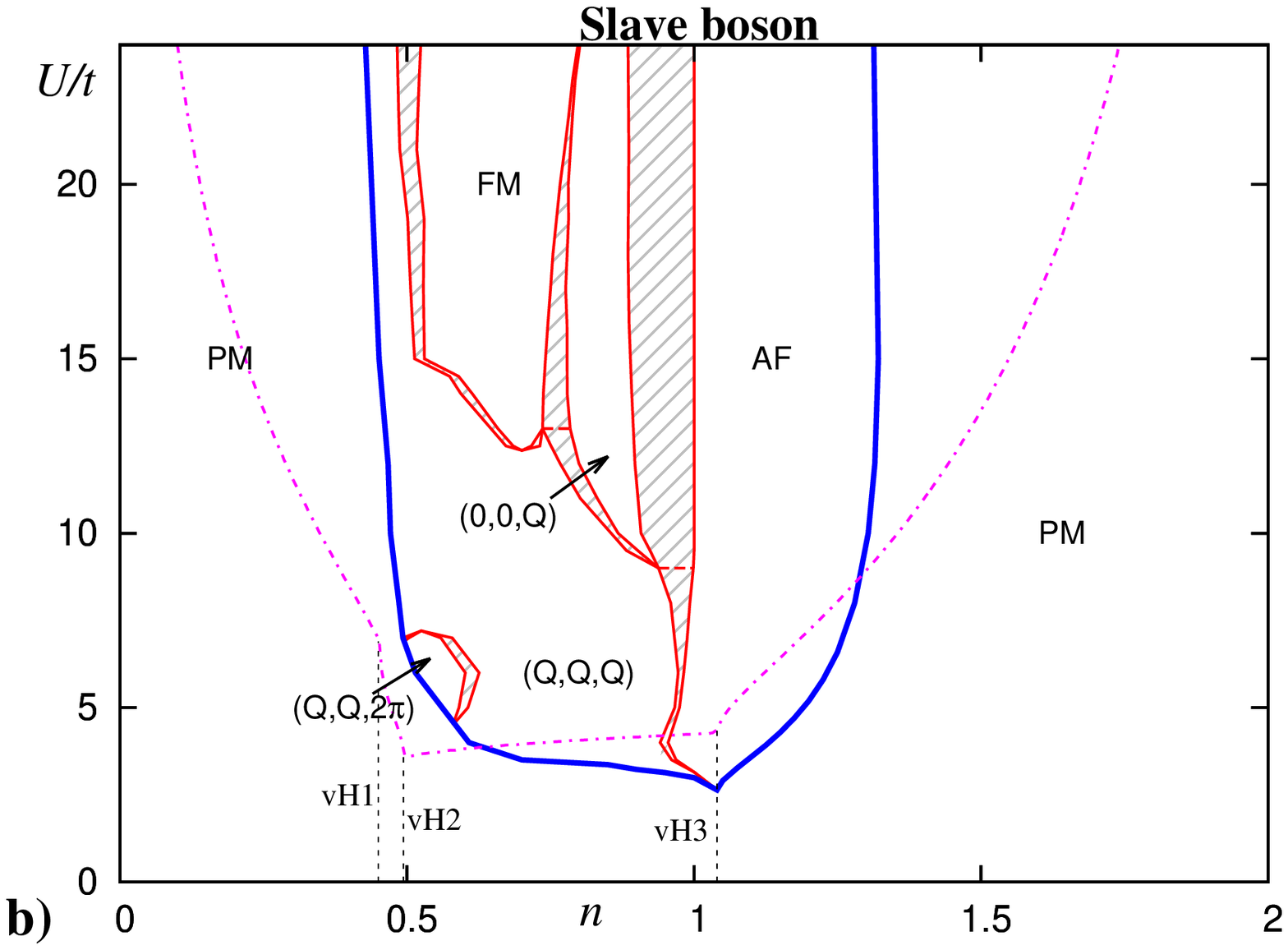}
\caption{
	(Color online)
	The same as in Fig. \ref{fig:square_t'=0.2} for the bcc lattice Hubbard model with $t'=0.3t$. (a) HFA results; (b) SBA results.
	$\mathbf{Q}_{\rm AFM} = (0,0,2\pi)$.
}
\label{fig:bcc_t'=0.3}
\end{figure}

We also present the results for the bcc lattice with $t'=0.3t$ obtained within HFA (Fig. \ref{fig:bcc_t'=0.3}a) and SBA
(Fig. \ref{fig:bcc_t'=0.3}b). In this case we have 3 bounded van Hove singularities: $n_{\rm vH1}\approx0.45$
(P point), $n_{\rm vH2}\approx0.494$ ($\kk=(\alpha,\alpha,\alpha)$, $\cos\alpha/2=t'/t$), $n_{\rm
vH3}\approx1.04$ (N point).
The variety of magnetic phases which exist far away from half-filling in HFA, vanishes.
The results are considerably different from those for the square and sc lattices. Despite the presence of van Hove
singularities at $n<1$ the FM ordering is strongly suppressed in favor of
spiral ordering. At moderate $U\gtrsim 12t$ the ground state FM ordering appears to be non-saturated ($1>m/n \geqslant0.5$),
and the saturated ferromagnetism occurs only at $U> 24t$.
For hole doping at rather large $U/t\gtrsim 10$ the FM phase and the (satellite) $\mathbf{Q}=(Q,Q,Q)$, $(0,0,Q)$
phases (they both can be considered as modulated FM phases) survive within SBA.
At smaller $U/t$ the diagonal $(Q,Q,Q)$ phase becomes favorable in a  wide region.

The AFM phase region is wide (especially in SBA) even away  from half-filling for $n>1$. 
The correlation effects stabilize pure AFM ordering
with wave vector $\mathbf{Q}=(0,0,2\pi)$ up to large $U/t$ in the vicinity of half-filling ($1<n\lesssim1.4$). This is explained by the nesting on the one hand and by the 'vH3' singularity on the other hand. $\mathbf{Q}=(0,0,2\pi)$ corresponds to the vector connecting $(0,\pi,\pi)$ and $(0,\pi,-\pi)$ N points of the Brillouin zone.
For
$n\gtrsim 1.4$ the AFM state changes to the PM phase through a second order transition.
The magnetic phase $\mathbf{Q}=(\pi,\pi,\pi)$ that occurs within HFA, as well as the corresponding PS region, turn out to be
 fully destroyed by the correlation effects.
Lowest critical values of $U$ appear to be practically not renormalized in the range between $n_{\rm vH2}$ and $n=1$,
otherwise the correlation effects destroy the magnetic ordering.

\subsection{Face-centered cubic lattice}
\label{sec:fcc}
\subsubsection{$t'=0$}
The results for non-bipartite lattices should be substantially different, which is implied by Nagaoka theorem for the
case of $U\rightarrow\infty$ \cite{Nagaoka66,Iordansky}.
\begin{figure}[h!]
\includegraphics[width=0.49\textwidth]{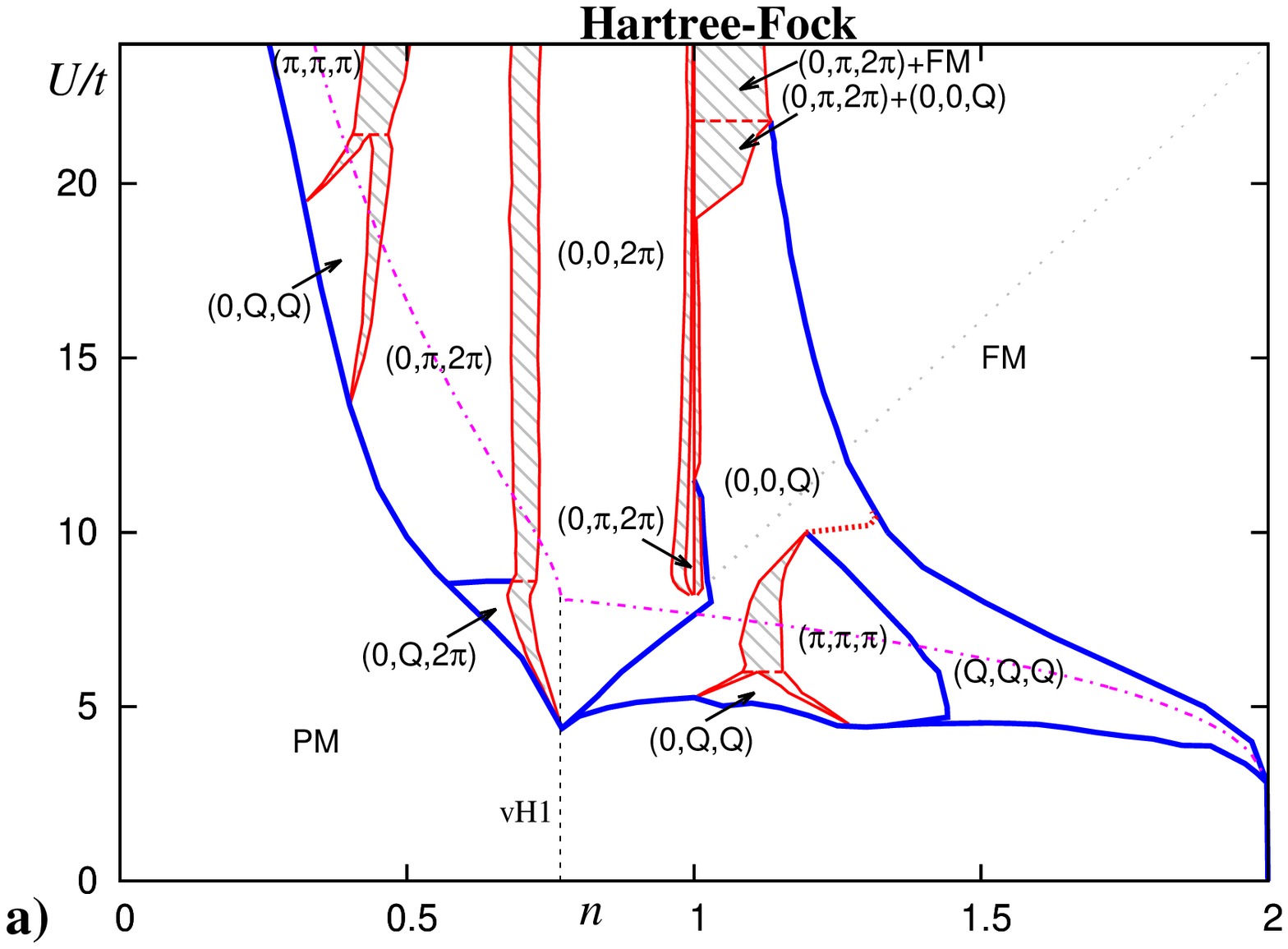}
\includegraphics[width=0.49\textwidth]{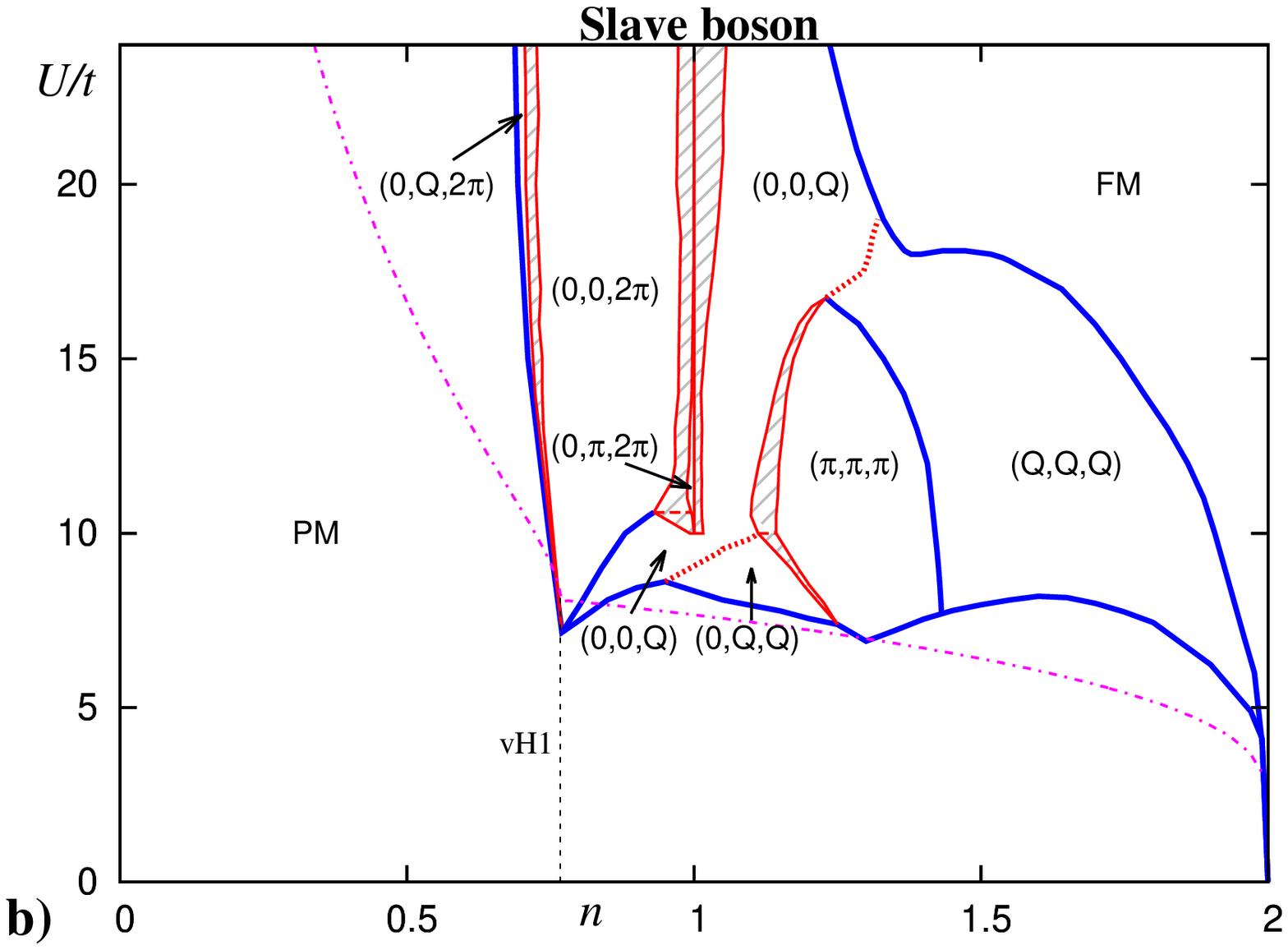}
\caption{
	(Color online)
	The same as in Fig. \ref{fig:square_t'=0.2} for the face-centered cubic lattice Hubbard model with $t'=0$ (a)
HFA results; (b) SBA results.
}
\label{fig:fcc_t'=0}
\end{figure}
In this section  we treat the fcc lattice.
Since this lattice is not bipartite, the symmetry with respect to particle--hole transformation breaks down.
Thus, for $t'=0$ we show separately the HFA  (Fig. \ref{fig:fcc_t'=0}a) and SBA (Fig. \ref{fig:fcc_t'=0}b) results for the density range $0<n<2$.
The bare density of states is characterized by a
divergent logarithmic singularity at the band top for $t'=0$ (originating from  the van Hove singularity line $\kk=(0,Q,2\pi)$, $0<Q<\pi/2$, V direction).
The important difference between the fcc lattice and those considered in the previous Sections is a good separation of the peculiar
regions: $\nVH=2$ (van Hove singularity at the band top) and $n=1$ (a very bad nesting) already at $t'=0$.

In the hole-doped half of the phase diagram we observe a strong
suppression of any magnetic ordering by correlations up to $n_{\rm vH1}\approx0.767$ which corresponds to the L point of the Brillouin zone. At larger density we obtain a quantitative renormalization of the lower phase boundaries with the PM phase. 
At half-filling we observe a very narrow region of antiferromagnetism corresponding to the wave vector
$\mathbf{Q}=(0,\pi,2\pi)$ (a non-collinear AFM structure)  which, however,
appears to be unstable with respect to competition with another type of commensurate antiferromagnetism (wave vector $\mathbf{Q}=(0,0,2\pi)$) at small hole doping
 and a vertically modulated FM phase (wave vector
$\mathbf{Q}=(0,0,Q)$) at small electron doping.
Another correlation consequence is a moderate suppression of the FM ordering in favor of the diagonal phase
(wave vector $\mathbf{Q}=(Q,Q,Q)$).
The FM ordering is saturated and appears to be stable well away from $n=2$ only at $U\gtrsim16t$.
In close vicinity of the band top the case of a small number of carriers is realized. Therefore, the
multiple scattering effects can substantially increase the critical value of $U$ for the PM--FM transition: probably the
most suitable way for the treatment of this problem is the $T$-matrix approach by Kanamori \cite{Kanamori1963}.
We believe that away from the band top (bottom) the slave-boson approximation yields a reasonable picture of magnetic
ordering.  The case of large $U$ will be treated in detail elsewhere.
\begin{figure}[h!]
\includegraphics[width=0.49\textwidth]{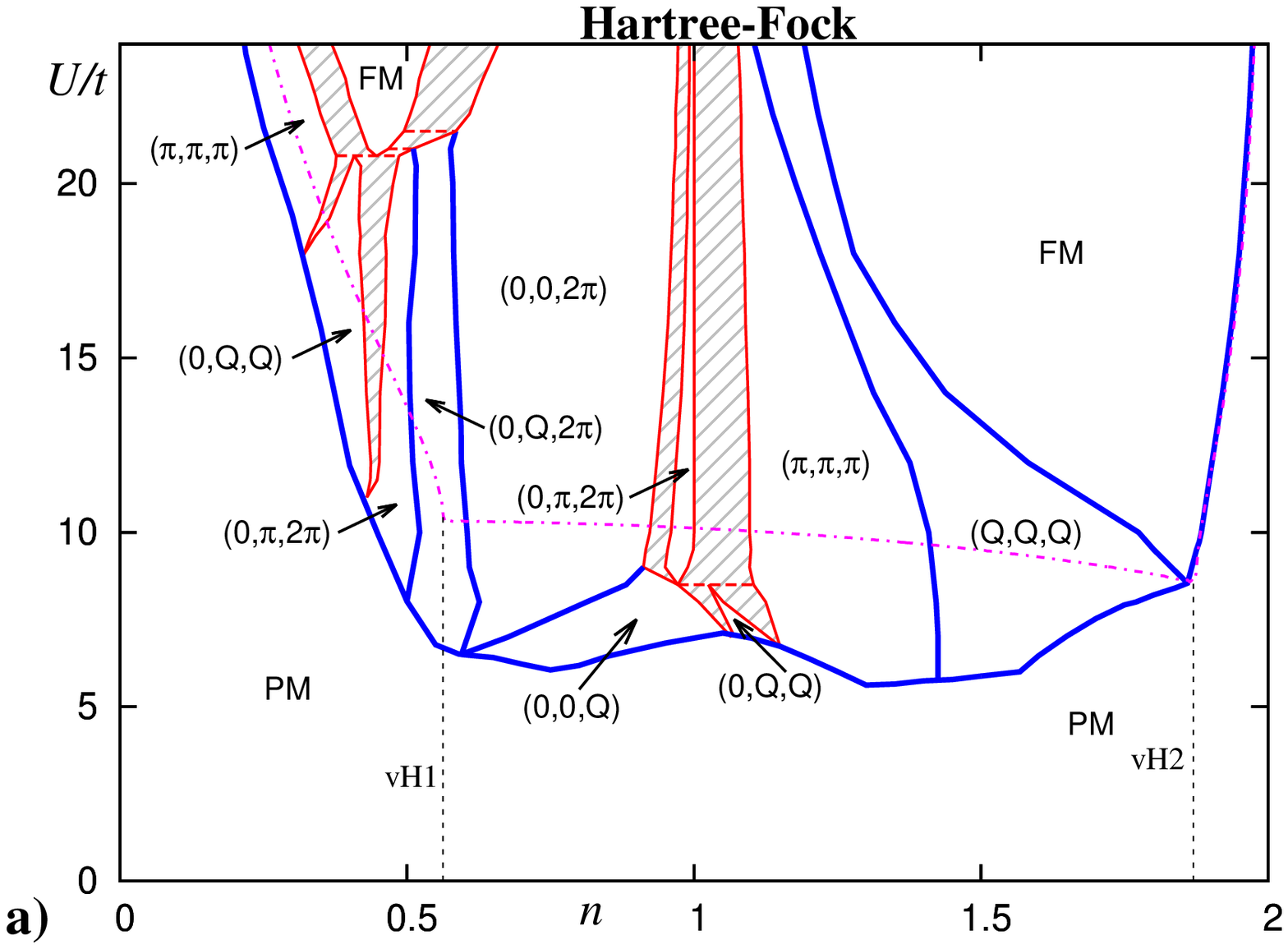}
\includegraphics[width=0.49\textwidth]{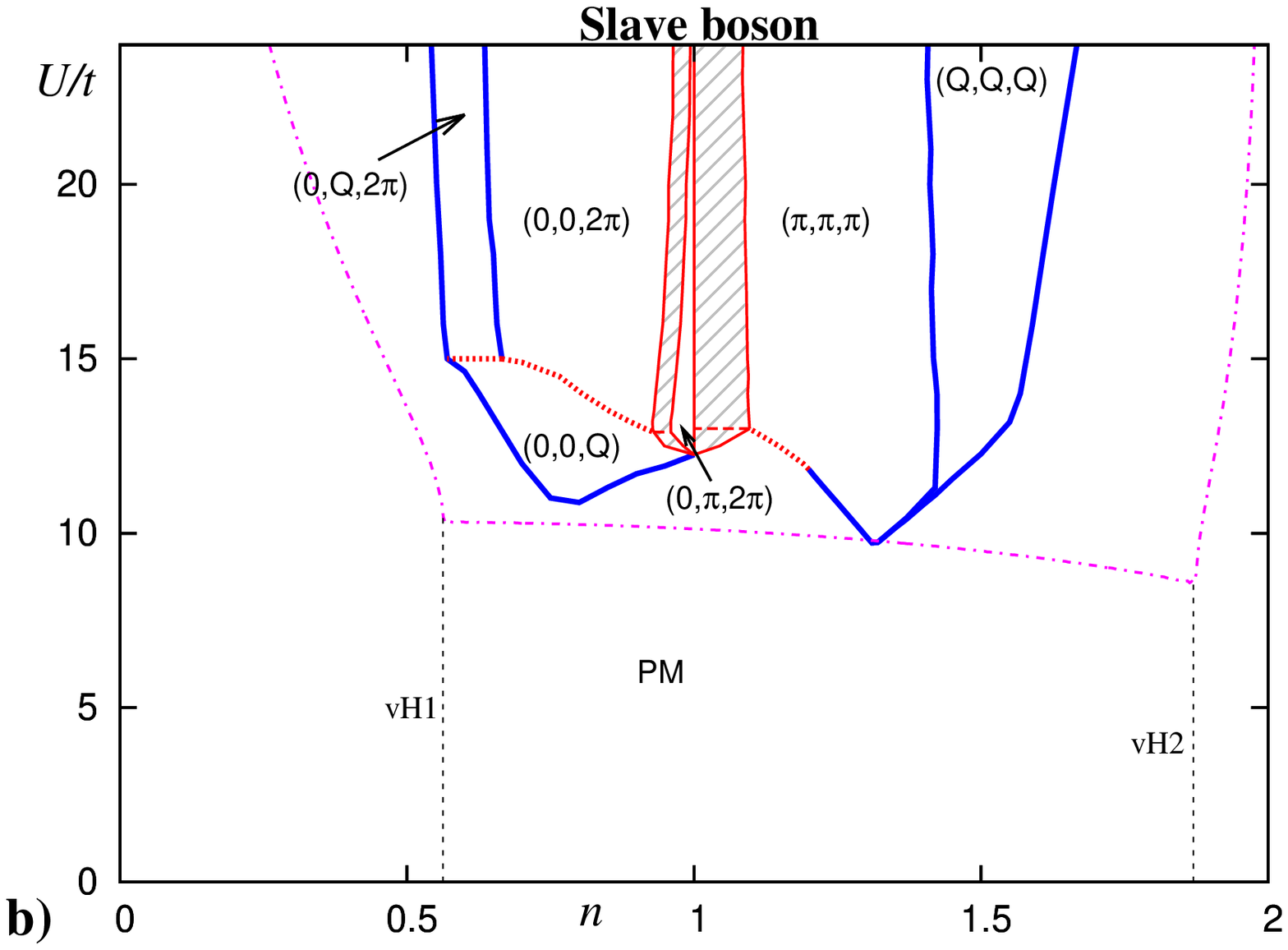}
\caption{
	(Color online)
	The same as in Fig. \ref{fig:square_t'=0.2} for the face-centered cubic lattice Hubbard model with $t'=0.3t$.
}
\label{fig:fcc_t'=+0.3}
\end{figure}

\begin{figure}[h!]
\includegraphics[width=0.49\textwidth]{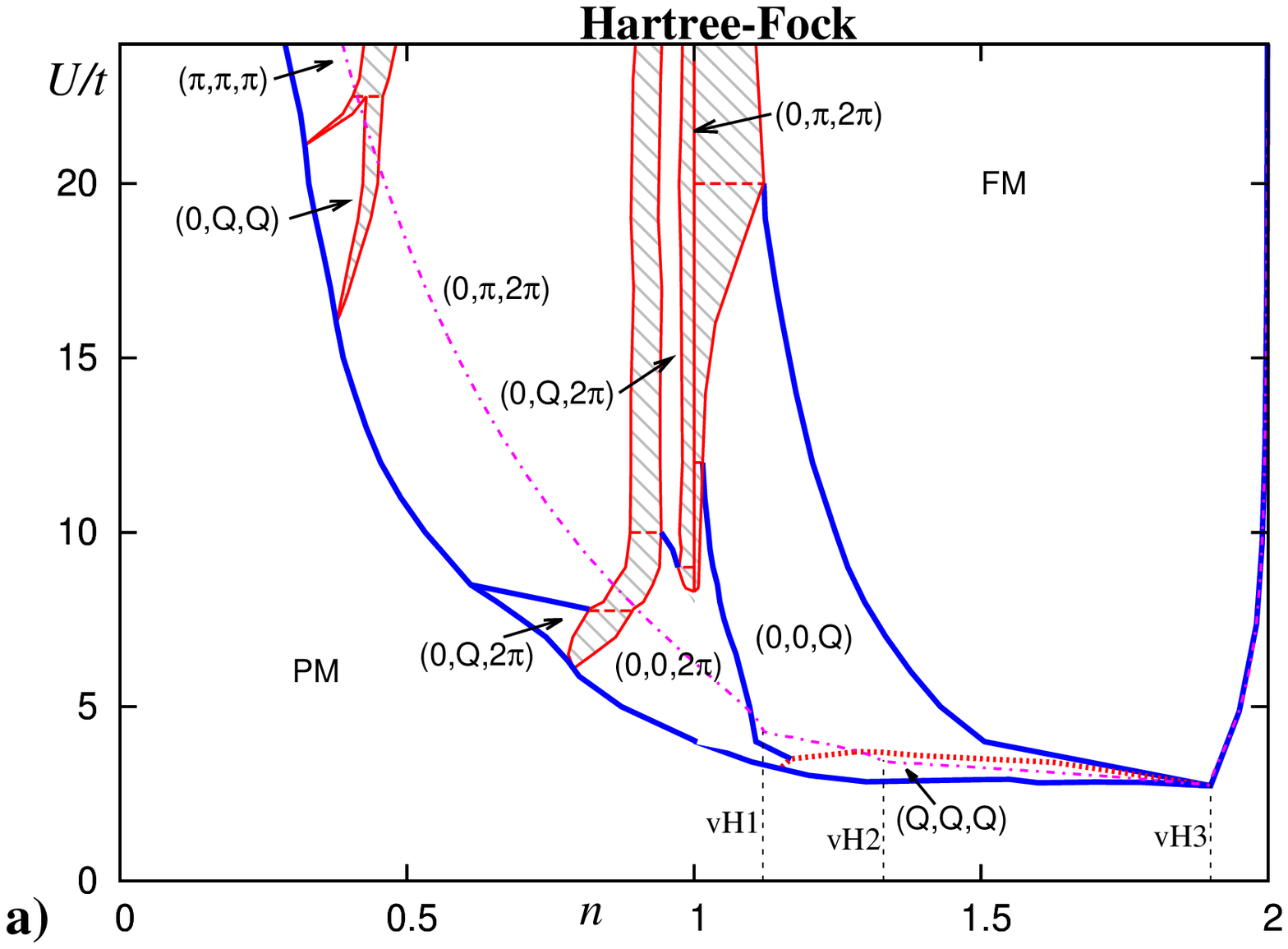}
\includegraphics[width=0.49\textwidth]{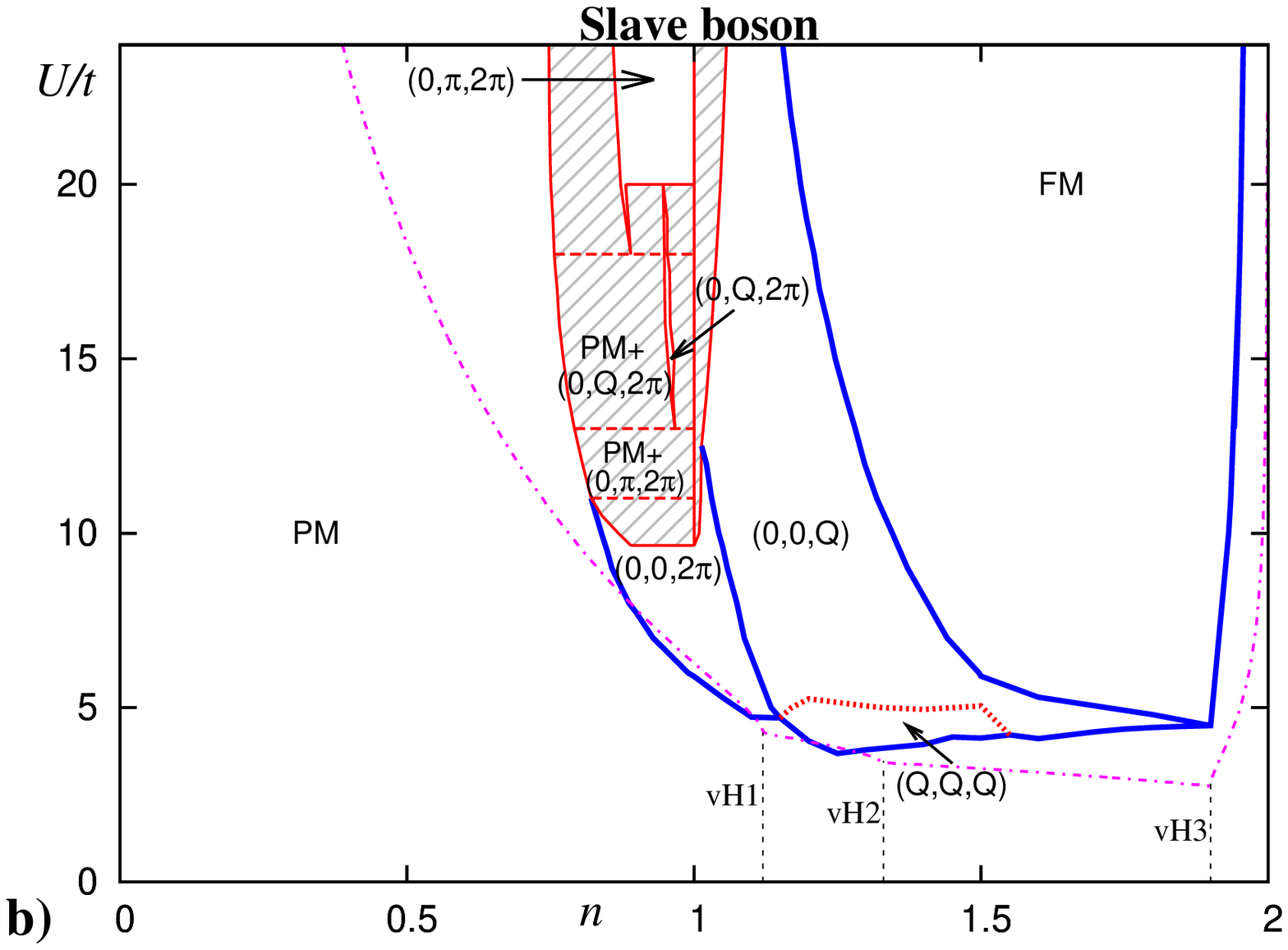}
\caption{
	(Color online)
	The same as in Fig. \ref{fig:square_t'=0.2} for the face-centered cubic lattice Hubbard model with
$t'=-0.3t$.
}
\label{fig:fcc_t'=-0.3}
\end{figure}

To trace the impact of finite $t'$ (which implies a finite DOS for the fcc lattice due to the cutting of the
logarithmic singularity), we calculated the ground state magnetic phase diagram in both HFA and SBA for $t'=0.3t$
(Fig.\ref{fig:fcc_t'=+0.3}) and $t'=-0.3t$ (Fig. \ref{fig:fcc_t'=-0.3}).

\subsubsection{$t'=0.3t$}

It is remarkable that for the case $t'=0.3t$ the correlation effects dramatically rebuild the phase diagram fully
destroying any magnetic ordering in the vicinity of the band top and replacing the saturated FM ordering by the diagonal
phase regions ($\mathbf{Q} = (Q,Q,Q)$ and $(\pi,\pi,\pi)$) well away from it.
The spiral and AFM phase regions are substantially narrowed far away from the value
$n=1$. There are 2 van Hove singularities in this case: $n_{\rm vH1}\approx0.563$ (L point), $n_{\rm
vH2}\approx1.87$ (W point). Both singularities correspond to high symmetry points of the Brillouin zone, which seems to
explain a rather small DOS for all the range of electron concentrations. Together with the absence of Fermi surface
nesting this leads to a particularly strong suppression of magnetism by electron correlations as compared to all the diagrams
studied.

\subsubsection{$t'=-0.3t$}

On the other hand, at $t'=-0.3t$ there appears an additional saddle point  with reduced symmetry $n_{\rm
vH3}\approx1.9$ ($(0,\alpha,\alpha)$, $\cos\alpha/2=t/(2t'-t)$) besides two other singularities $n_{\rm
vH1}\approx1.12$ (L point), $n_{\rm vH2}\approx1.33$ (X point). The 'vH3' singularity leads to a relatively high
DOS and consequently to low values of $U_c$. As a result,
renormalizations of the
phase diagram in the region between $n_{\rm vH3}$ and $n=1$ become rather small.
Above $n_{\rm vH3}$ the FM ordering is strongly suppressed, which dramatically disagrees with the Stoner
criterion and HFA calculations, but agrees with the $T$-matrix approximation results \cite{Kanamori1963}.
However, when the number of carriers becomes equal to or less than $n_{\rm vH3}$ we find a large
saturated FM ordering region even at small
$U$, which is suppressed by the spiral with $\mathbf{Q}=(0,0,Q)$ in the vicinity of half-filling (this picture holds
also
within HFA). In general we conclude that the magnetic ordering in the electron-doped half of the phase diagram is only
quantitatively affected by the correlations: even the Stoner criterion works fine in this region. In contrast to the other lattices the van Hove scenario does not determine wave vector in vicinity of $n_{\rm vH3}$'s in general case, the high DOS factor prevails in the electron-doped half and causes a tendency to structures with 
small wave vectors.
In the hole-doped half of the phase diagram we find the typical for  SBA suppression of any magnetic order far away
from $n=1$. What is unusual in the hole-doped half, is the presence of wide regions of PS between the PM and
different spiral magnetic states.

\subsection{Analysis of bosonic fields: comparison of Hartree--Fock and slave-boson approximations}
\label{sect:analysis}

\begin{figure}[h!]
\includegraphics[width=0.49\textwidth]{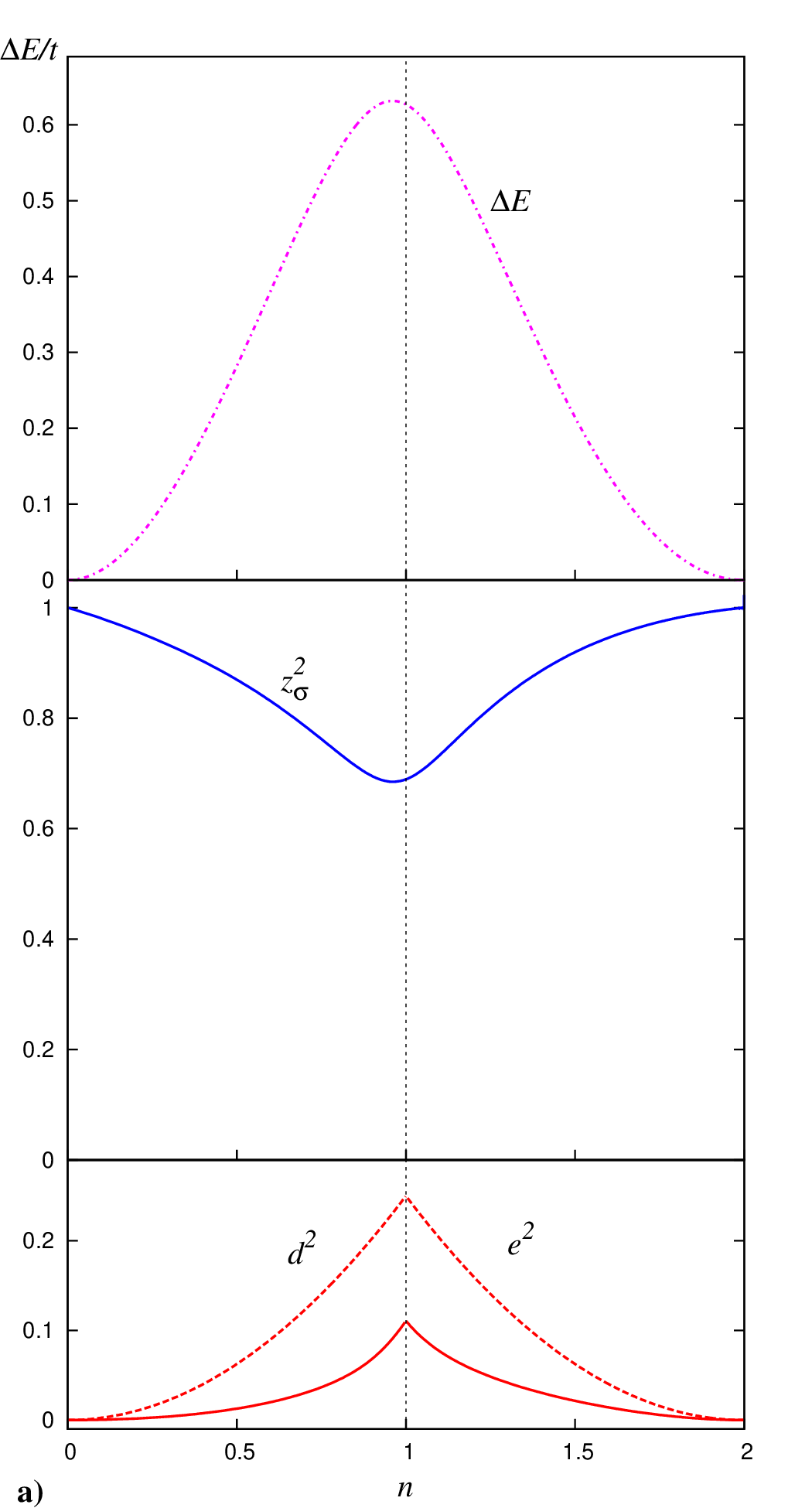}
\includegraphics[width=0.49\textwidth]{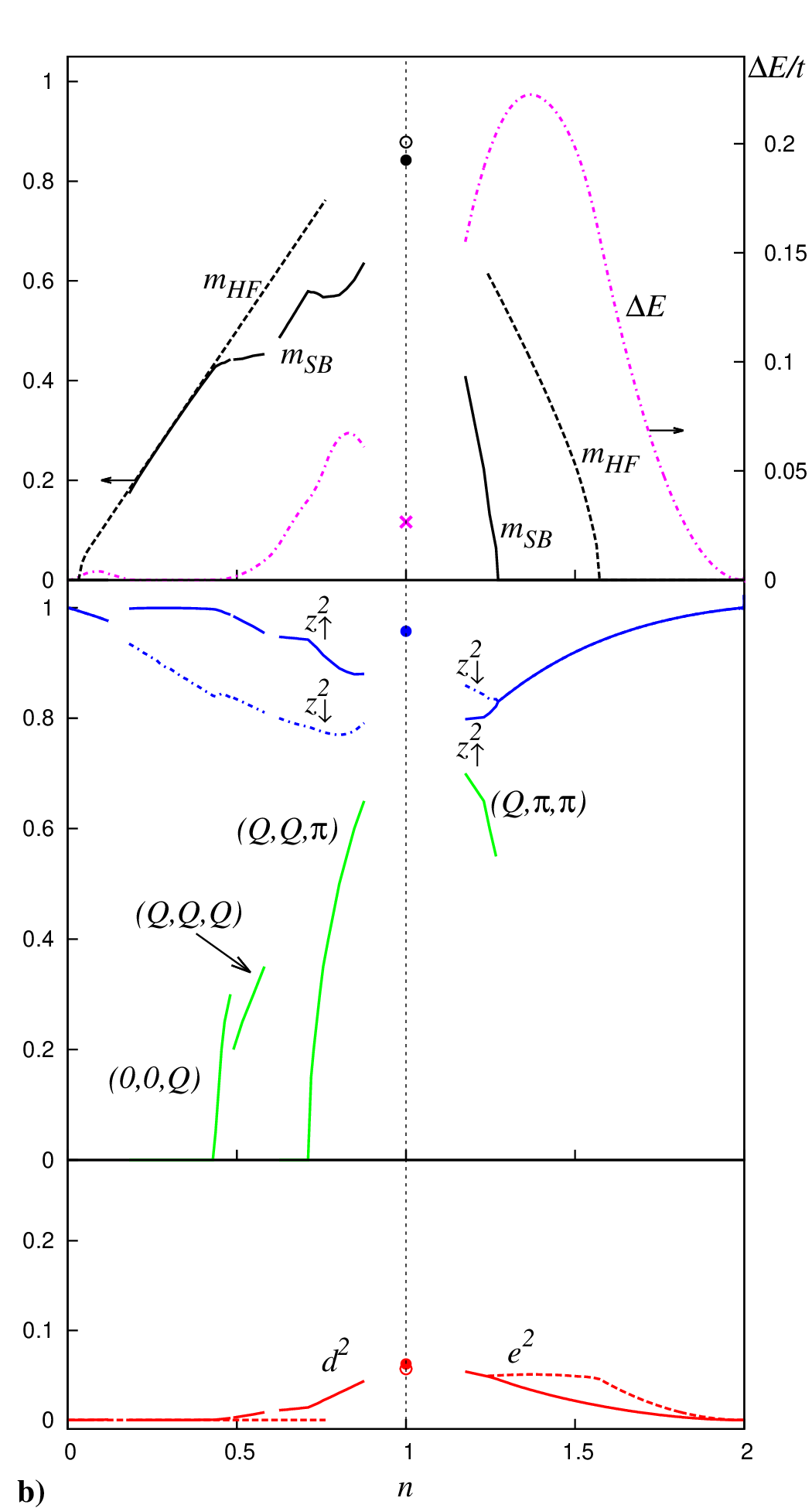}
	\caption{(Color online) The density dependence of the physical quantities for simple cubic lattice with $t'=0.3t$ and $U = 9t$  for (a) paramagnetic state and (b) magnetically ordered ground state. Upper panel:
	The magnetic moment $m$ calculated within SBA (HFA) is shown by black solid (dashed) line on left axis. 		Filled (empty) circle shows the magnetization for AFM phase at $n=1$ within SBA (HFA). 
		The energy difference between HFA and SBA $\Delta E=E_{\rm HF}-E_{\rm SB}$ is shown by purple dash--dotted line on right axis. The cross shows this energy difference for AFM phase at $n=1$.     
	Middle panel:  The renormalization factor $z^2_{\uparrow}$ ($z^2_{\downarrow}$) is shown by blue solid  (dashed) line; the circle shows it for AFM phase at $n=1$. $Q_{x,z}/\pi$ is shown by green solid line.
	Lower panel: Occupancies of the double carrier ($d^2$ for $n<1$ and $e^2$ for
	$n>1$) calculated within SBA (HFA) are shown by solid (dashed) red line; filled (empty) circle shows it for AFM phase at $n=1$ calculated within SBA (HFA). 
}
\label{fig:bosonic_vars}
\end{figure}



To illustrate the correlation effects impact in detail, we show the concentration dependence of site occupancies for the
forced PM state and the magnetically ordered state within both HFA (given by Eq. (\ref{eq:d_eq})) and SBA (Eq. (\ref{eq:main_p})).
Figure \ref{fig:bosonic_vars} depicts the behavior of bosonic variables depending on the
electron concentration $n$ for the sc lattice with $t'=0.3t, U=9t$   for the PM phase (a) and the
spiral (antiferromagnetic) ground state (b).
We present $d^2$ for $n<1$ and $e^2$ for $n>1$, since doubles and holes play the role of current carriers.
Below we discuss for brevity the case of $n<1$, analogous conclusions hold for $n>1$ with the replacement $d\rightarrow e$.
In the PM case $d^2$ highly increases as $n$ approaches the half-filling within HFA, which strongly disagrees
with its suppression (and the corresponding bandwidth narrowing $z^2$) within SBA.
The latter results in the metal--insulator transition for $n=1$ at a sufficiently large $U/t$ \cite{Kotliar86} when the spectral weight $z^2$ vanishes.
This treatment of the metal--insulator transition is physically close to the Brinkman--Rice one \cite{Brinkman}.
Therefore, within SBA, in the PM state electrons avoid double site occupancy, which allows their Coulomb energy to be
lowered (the price is an increase in kinetic energy due to the band narrowing):
in bosonic language this means a much more favorable way of redistribution between $p^2$ and $d^2$ than in HFA ($d_{\rm SBA}\ll d_{\rm HFA}$).

A very different picture can be found for the magnetically ordered case, where the system has an additional degree of freedom (redistribution between $p^2_\uparrow$ and $p^2_\downarrow$).
We found that already within HFA the Coulomb energy turns out to be substantially reduced due to the lowering of $d^2$, and the distinction from the SBA in this sense is small.
In this case one finds that the main correlation effect is the redistribution between $p^2_\uparrow$ and $p^2_\downarrow$ which results in:
(i) a partial magnetization suppression, (ii) a moderate bandwidth reduction $z^2_\sigma$ due to the Pauli principle.
The general conclusion is that the correlation impact on the magnetically ordered state appears to be much more weaker
than on the PM phase. Therefore, the main ``correlation'' effect (in a broad sense) is the magnetism formation itself. 
The data on magnetic moment and $z$-factors of Fig.~\ref{fig:bosonic_vars}b demonstrate the jump of magnetic moment near first order transition ($\Delta m \approx 0.2$ for $n<1$ and $\Delta m \approx 0.4$ for $n>1$). The renormalizations of effective masses given by $z^{-2}_\sigma$ are considerably spin-dependent. In particular, at large $U$ in the ferromagnetic region we have $z_{\uparrow}^2 \simeq 1$ and $z_{\downarrow}^2 \simeq 1 - n = e^2$. 
In fact, in this limit the spin-down states have non-quasiparticle nature\cite{Katsnelson}.

For the half-filling $e^2$ and $d^2$ are equal, being small (Fig. \ref{fig:bosonic_vars}b), which implies that the
spectra
renormalization factor $z^2$ tends to $1$ at $n=1$. 
The increase of $U$ causes the change of the gap nature from the Slater to Hubbard one, the latter being retained in the paramagnetic state. In particular, the difference between mass renormalization within ordered and paramagnetic case was discussed in Ref.~\onlinecite{Korbel}. However, near half-filling HF works well for the electron and spin-wave spectrum at  $U \gg |t|$, yielding exactly atomic limit\cite{Entelis}. 
While in the ordered state the mass enhancement, $1/z_{\sigma}^2$, is smallest at $n = 1$, at doping  this quantity rapidly grows as one can see from Fig.~\ref{fig:bosonic_vars}b. 
This explains why the SBA results are very
similar to the HFA ones in the vicinity of half-filling: the sequences of the
magnetic states on the diagrams do not change
with increasing $U$, although their regions shift to larger $U$'s in SBA (Fig. \ref{fig:bosonic_vars}b).

Upper panels of Fig. \ref{fig:bosonic_vars} demonstrates also the energy difference $\Delta E=E_{\rm HF}-E_{\rm SB}$ between HFA and SBA. $\Delta E$ for PM phase reaches $0.63t$ and is much larger than $\Delta E$ for magnetic state, which reaches $0.23t$ only. This illustrates also the conclusion that correlation effects lower the energy of PM phase much stronger than the energy of magnetic states. The energy lowering at half-filling is maximal for PM phase and minimal for spiral state, which agrees with the dependencies of $z$-factors.

We note that for other lattices the results for bosonic fields and the energy difference behavior are qualitatively the same.

\section{Conclusions}
\label{sec:conclusions}
We have investigated the ground state magnetic phase diagram including spiral ordering formation and phase separation within both Hartree--Fock and slave--boson approaches. 
The slave-boson approximation enables one to take account of the correlations in terms of a few parameters
renormalizing
the electron spectrum.
The main advantage of SBA is a correct qualitative description of the paramagnetic phase: the energy
becomes lower as compared with HFA, so that the gain in energy  of the magnetic phases is now substantially smaller. As a
result, the paramagnetic region occurs in the ground state within the SB method at large $U$.
Being technically simpler in comparison with DMFT,  SB mean--field approximation gives a transparent picture of electron spectrum, but misses dynamical self-energy effects. 
However these effects can be in principle taken into account within SBA beyond mean--field level. 

Spiral configurations within DMFT for the square lattice with $t'=0$ were treated in Refs.~\onlinecite{Fleck1998,Fleck1999}. 
The sequence of phase transitions for $U=8t$ is AF$\rightarrow$ AF$+(Q,Q)\rightarrow(Q,Q)\rightarrow(Q,Q)+(Q,\pi)\rightarrow(Q,\pi)\rightarrow$ PM. This is in agreement with out results, although the corresponding $U$ value is somewhat larger. According to Ref.~\onlinecite{Fleck1998}, magnetic ordering vanishes for $n<n_{\rm c} = 0.35$, whereas our critical value $n_{\rm c}$ is considerably larger, about 0.6.  
Recent inhomogeneous DMFT calculations \cite{Peters2014} enabled one to describe formation of spin density wave and stripe states in the 2D case. 
The results are in a qualitative agreement with our data (see Fig. \ref{fig:square_t'=0}). 
A quantitative difference of the critical concentration ($n_{\rm c}=0.8$ instead of our value 0.6 for $U\gtrsim10t$) may be due to that we consider spirals rather than spin density waves.
In Ref.~\onlinecite{Oles} a comparison of spiral and stripe states was performed within the SB method for different $n$ and $t'/t$. The spirals turn out to be more energetically favorable for large $t'/t $ and $n=7/8$. 
Besides that, stripe states may  are spatially homogeneous and loose in long--range Coulomb interaction energy in comparison with spiral states.

Our results and their comparison with DMFT enable one to estimate the role of dynamical corrections to the electronic spectrum, since the SBA treats the nature of correlated electron states only on the static level. 

The results obtained for both 2D and 3D cases can be applied to an analysis of the properties and phenomena in $d$-metals and their compounds,
especially with fcc and bcc structures. The correlation effects considered can be important, e.g.~for the metal--insulator (Mott) transition (this problem was treated in Ref.~\onlinecite{Timirgazin12} in the Hartree--Fock approximation including the spiral states).

We see that the correlation effects lead to a strong suppression of the regions of the existence
of the magnetic phases.
At the same time, there remain the first-order transitions and  noticeable regions of the phase separation between the
magnetically ordered states.
The correlation effects near half-filling change only slightly the Hartree--Fock results, so that at small $t'/t$
the sequences of magnetic states do not change with increasing $U$.

Both SBA and HFA yield qualitatively (but not quantitatively) similar results near half-filling, including phase separation.  
The main advantage of SB method with respect to HFA is correct account of the doubly occupied sites onto the one-electron spectrum which manifests itself through the $z$-factors. 
Owing to them PS regions become more narrow within SB method. 

Our results confirm the importance of the van Hove and nesting scenarios in stabilization of spiral magnetic structures. 
The divergent van Hove singularities which are characteristic for 2D lattices, as well as for bcc  lattice with $t'=0$, prevent the suppression of magnetism by the electron correlations. 
The role of spectrum renormalization within SBA in comparison to HFA is weaker when the physics is governed by van Hove singularities. 
Far from half-filling, the
van Hove singularities stabilize the ferromagnetic order, especially in the 2D case. The correlation effects are shown
to be negligible for the FM state in the case of a large density of states, and for the spiral states provided that the Fermi surface
nesting with corresponding ${\bf Q}$  is present. The most important among the bounded singularities are those  produced by the saddle points which do not correspond to high symmetry points of the Brillouin zone. They favor 
stabilization of incommensurate spiral states and usually lead to an increase in the density of states. The value of
wave vector ${\bf Q}$ is equal to a vector connecting two of the saddle points  in these cases.
When the system is far from half-filling and the Fermi level is far from van Hove singularities, the magnetic state cannot form at any $U$.


A recent series of papers \cite{Kugel2013_pniktides,Kugel2013_two-bands} concerns the formation of spiral magnetic
order in a specific many-band model (two-band Rice model~\cite{Rice1970}) with imperfect nesting between the hole and
electron Fermi surface pockets.  
The general result of these investigations is that there exists a first order transition between commensurate and incommensurate magnetic states which invokes a phase separation between them.  
Thus the phase separation is a more general phenomenon and the existence of two nested Fermi surfaces is not necessary. 

\section{Acknowledgements}
\label{sec:akhnoledgments}
We are grateful to A.A.~Katanin, A.O.~Anokhin, A.I.~Lichtenstein and K.I.~Kugel for useful discussions. 
The research was carried out within the state assignment of FASO of Russia (theme ``Electron'' No. 01201463326). 
This work was supported in part by the Division of Physical Sciences and Ural Branch, Russian Academy of Sciences
(project no. 15-8-2-9, 15-8-2-12) and by the Russian Foundation for Basic Research (project no.
14-02-31603-mol-a) and Act 211 Government of the Russian Federation 02.A03.21.0006. 
The main amount of calculations was performed using the ``Uran'' cluster of IMM UB RAS.

\end{document}